\documentclass{aa}  
\usepackage{graphicx}
\usepackage{txfonts}
\usepackage{xcolor}
\usepackage{url}
\usepackage{hyperref}
\usepackage{natbib}
\hypersetup{
    colorlinks=true,
    breaklinks=true,
    urlcolor= red,
    linkcolor= blue,
    citecolor= blue
    }
\definecolor{BlueF}{rgb}{0.03, 0.27, 0.49}

\definecolor{ochre}{rgb}{0.8, 0.47, 0.13}

\definecolor{fuchsia}{rgb}{0.57, 0.36, 0.51}

\begin{document} 
    \title{Dynamical effects of the radiative stellar feedback on the H\,{\textsc i}-to-H$_2$ transition}
    \authorrunning{Vincent Maillard et al.}
    \titlerunning{Dynamical effects of the radiative stellar feedback on the H\,{\textsc i}-to-H$_2$ transition}
    \author{Vincent Maillard\inst{1}, Emeric Bron\inst{1}, \and Franck Le Petit\inst{1}}
    \institute{LERMA, Observatoire de Paris, PSL University, CNRS, Sorbonne Université, 92190 Meudon, France\\
              \email{vincent.maillard@observatoiredeparis.psl.eu}}
    \date{Received 23 March 2021 / Accepted 11 September 2021}
    \abstract
        {The atomic-to-molecular hydrogen (H/H$_2$) transition has been extensively studied as it controls the fraction of gas in a molecular state in an interstellar cloud. This fraction is linked to star-formation by the Schmidt-Kennicutt law. While theoretical estimates of the column density of the H\,\textsc{i} layer have been proposed for static photodissociation regions (PDRs), Herschel and well-resolved ALMA (Atacama Large Millimeter Array) observations have revealed dynamical effects in star forming regions, caused by the process of photoevaporation.}
        {We extend the analytic study of the H/H$_2$ transition to include the effects of the propagation of the ionization front, in particular in the presence of photoevaporation at the walls of blister H\,\textsc{ii} regions, and we find its consequences on the total atomic hydrogen column density at the surface of clouds in the presence of an ultraviolet field, and on the properties of the H/H$_2$ transition.}
        {We solved semi-analytically the differential equation giving the H$_2$ column density profile by taking into account H$_2$ formation on grains, H$_2$ photodissociation, and the ionization front propagation dynamics modeled as advection of the gas through the ionization front.}
        {Taking this advection into account reduces the width of the atomic region compared to static models. The atomic region may disappear if the ionization front velocity exceeds a certain value, leading the H/H$_2$ transition and the ionization front to merge. For both dissociated and merged configurations, we provide analytical expressions to determine the total H\,{\textsc i} column density. Our results take the metallicity into account. Finally, we compared our results to observations of PDRs illuminated by O-stars, for which we conclude that the dynamical effects are strong, especially for low-excitation PDRs.}
        {}
    \maketitle
    \section{\label{sec_introduction}Introduction}
        In the atomic envelop surrounding a molecular cloud \citep{Wannier1983, Andersson1991}, the transition where molecular hydrogen turns into atomic hydrogen is called the atomic-to-molecular hydrogen (H/H$_2$) transition, or the photodissociation front. Knowledge about the location of the H/H$_2$ transition gives us information on the total fraction of molecular gas in a cloud. It is indeed in the cold and dense molecular medium that stars are formed through gravitational collapse of the gas. The H$_2$ mass fraction is linked to star-formation thresholds by the Schmidt-Kennicutt relations (\citealt{Schmidt59, Kennicutt98}, see also \citealt{Tacconi2020} for a review on molecular gas mass in galaxies and its link to star formation rates and the evolution of galaxies). Secondly, the intensity of the lines of the molecular tracers depends on the depth of the H/H$_2$ transition due to the specific chemistry occurring in the warm H$_2$ gas.
        
        Estimating the total H\,{\textsc i} column density at the surface of neutral interstellar clouds is a key problem for tackling several important astrophysical tasks:
        to deduce the H$_2$ mass fraction intervening in the Schmidt-Kennicutt law of star-formation \citep{Villanueva17}, to deduce the CO-to-H$_2$ conversion factor \citep{Schruba17, Dessauges17, Pineda17}, or even to study the dynamics of a cloud \citep{Valdivia16, Beuther20}. Some direct estimates of the total H\,{\textsc i} column density were also performed through 21 cm observations \citep{Kim1998, Stanimirovic1999, Stanimirovic2014, Kim2003, Staveley-Smith2003, Stil2006, Peek2011, Bihr2015, Beuther2016}.

       A first analytical model of the H/H$_2$ transition was presented by \cite{Sternberg88} who derived an analytic formula for the total H\,{\textsc i} column density for a cloud with planar geometry illuminated by a beamed radiation field. He provided results as a function of the far-ultraviolet (FUV) radiation intensity, the cloud gas density, and the metallicity. In two papers, \cite{Krumholz08, Krumholz09} presented new models for the H/H$_2$ transition taking into account the metallicity dependence of the molecular mass fraction in galaxy disks and multidimensional geometry of finite-sized clouds embedded in ambient isotropic fields. More recently, \cite{Sternberg14} and \cite{Bialy2016} updated the analytic theory for one-dimensional (1D) planar slabs and spheres for beamed and isotropic radiation fields, aiming to apply them to global galaxy evolution studies. Their results were compared and validated by Meudon PDR Code \citep{LePetit2006} computations. These models are widely used to predict the column density of the atomic surface layer \citep{Wong2009, Bolatto2011, Lee2012, Lee2015, Roman-Duval2014, Bialy15, Bialy2017, Schruba2018, Noterdaeme2019, Klimenko2020a}. They have been extended to other problems such as the determination of the C$^+$/H$_2$ ratio \citep{Nordon2016} or the HD/H$_2$ ratio in the diffuse interstellar gas \citep{Balashev2020}. Also, \cite{Bialy2017b} explored the effects of turbulence on the H/H$_2$ transition using analytic and numerical methods with an application to the Perseus molecular cloud. Recently, \cite{Sternberg2021} extended the theory to the H/H$_2$ transition in the dust-free primordial interstellar gas.
       
        In the analytic theories presented above, the medium was assumed to be in a static and stationary state. However, clues about dynamical effects were recently found. Herschel observations of excited lines in strongly UV-illuminated galactic photodissociation regions (PDRs) \citep{Joblin18, Wu18} revealed a thin compressed surface layer ($10^{-3}~\mathrm{pc}$) with a thermal pressure well above its environment at the edge of the PDR ($P_{\mathrm{th}}\sim10^{8} \, \mathrm{K \, cm^{-3}}$). The gas thermal pressure was found to be strongly correlated to the UV field intensity.
        Moreover, Atacama Large Millimeter Array (ALMA) observations \citep{Goicoechea16, Goicoechea17} revealed the spatial structure of the Orion Bar PDR with an excellent resolution, showing that hot molecular tracers were emitting from a thin layer at high values of pressure and temperature. It was proposed that photoevaporation at the ionization front could explain the high pressures found in the neutral PDR and their correlation between the thermal pressure and the UV field intensity. Numerical models of photoevaporating PDRs with the time-dependent and dynamical Hydra PDR Code \citep{Bron18} reproduced the pressure structure and the correlation of the pressure with the UV radiation field.
   
        When a star illuminates a neutral cloud, its surface is heated by the radiation field. If the pressure in the H\,\textsc{ii} region is not sufficient to contain the heated gas (e.g., dense globules embedded in the H\,\textsc{ii} region, \citealt{Bertoldi1989} and \citealt{Bertoldi96}, blister H\,\textsc{ii} regions where the ionized gas is free to escape to the surrounding diffuse medium, \citealt{Israel1978b} and \citealt{Tenorio-Tagle1979A}), a photoevaporation flow of ionized gas streams from the neutral cloud into the H\,\textsc{ii} region. The theory of ionization fronts was first derived by \cite{Kahn1954} following the suggestion of \cite{Oort1955}, and is presented in textbooks (e.g., \citealt{Spitzer1978}, Chapter 12.1, \citealt{Draine2011}, Chapter 37). This theory shows that steady ionization fronts cannot exist for a specific range of ionizing photon flux to neutral gas density ratios, $F_\mathrm{EUV}/n_\mathrm{H}$, which thus separates two types of allowed solutions, called R-type and D-type. A R-type ionization front propagates supersonically relative to the neutral gas, and can occur either during the initial phase of development of an H\,\textsc{ii} region or in a later phase when encountering a steep decreasing density ramp \citep{Franco1989, Franco1990}. A D-type front propagates subsonically relative to the neutral gas, the gas is accelerated when passing through the ionization front and a pressure jump is maintained between the ionized and the neutral gas. When conditions fall in the forbidden range between R-type and D-type, a (usually low-velocity) shock propagates ahead of the ionization front in the neutral gas, raising the neutral gas density to reach D-type conditions. After an initial phase, this shock front overtakes the H/H$_2$ transition. Nearly critical D-type fronts are thought to be omnipresent in evolved, blister-type (\citealt{Israel1978b}, or "champagne flow", \citealt{Tenorio-Tagle1979A}) H\,\textsc{ii} regions found in many star forming regions and where archetypal dense PDRs are found (e.g., the Orion Bar in the Orion Nebula, \citealt{Odell2001}). Advection of the neutral gas from the neutral PDR region through these D-type fronts can thus maintain the H/H$_2$ chemistry out of equilibrium.
        
        This photoevaporation mechanism, proposed to explain the dynamics of bright PDRs illuminated by young stars at the edges of molecular clouds bordering blister H\,\textsc{ii} regions, has been studied in the past to describe the outer layer of small neutral molecular globules embedded in compact H\,\textsc{ii} regions by \cite{Bertoldi96} and \cite{Storzer98}. The upcoming James Webb Space Telescope (JWST) will give us access to high angular resolution observations of many H$_2$ rotational and ro-vibrational lines that will help probe the high pressure layer of PDRs and constrain the photoevaporation mechanism.
        
        In \citeyear{Bertoldi96}, \citeauthor{Bertoldi96} studied the PDR structure at the surface of a clump embedded in a H\,\textsc{ii} region by taking into account the expansion of newly ionized gas in the H\,\textsc{ii} region separating the cloud from one or more O stars. Thanks to the curvature of the modeled clump, the gas can expand freely, so that the ionizing radiation erodes the gas. This translates into a propagating ionization front, which is the transition between ionized hydrogen H$^+$ and atomic hydrogen. They linked the Extreme-UV (EUV, > 13.6 eV) and Far-UV (FUV, $\in \left[6-13.6 \right]$ eV) radiation flux arriving at the respective fronts (ionization and photodissociation) with their velocities. When the ionization front is slower than the dissociation front, an atomic region is developed, yet the velocity of the dissociation front is too high for stationary models of PDRs to still be applicable. In the other case, where the dissociation front is slower than the ionization front, the fronts are not separated and the atomic region does not exist. For a wide range of parameters, the atomic region is controlled by nonequilibrium effects, leading either to its nonexistence or to a smaller value of total H\,{\textsc i} column density than predicted with stationary models. 
   
        A following paper, \cite{Storzer98} presented a model of the thermal and chemical structure of the PDR including a propagating ionization front. They solved the full time-dependent structure with a constant ionization front velocity and a uniform density that does not vary with time. They obtained similar conclusion to those of \cite{Bertoldi96} about the merging of the fronts, and found that the H$_2$ rotational and vibrational line emission can be enhanced by a factor of 3.
        
        As of today, with Herschel, ALMA or upcoming JWST observations, dynamical effects can be highlighted in PDRs, leading to a need for an extension of \cite{Sternberg14} to include the effects of photoevaporation. In this paper we extend the H/H$_2$ transition model to take into account the dynamics induced by the photoevaporation mechanism in a new semi-analytical theory. As in \cite{Storzer98}, we model the propagation of the ionization front with a constant velocity, but find the structure of the PDR by solving the stationary equation of the chemistry including advection. This equation depends only on the physical conditions that are the velocity of the ionization front, the FUV field intensity, the gas density, and metallicity.
        
        Our goal is to compute the impact of the dynamical effects on the position and width of the H/H$_2$ transition, and its consequences on total atomic hydrogen column density. In Sect. \ref{sec_analytic_overview} we discuss our semi-analytical approach to solve the equation driving the H$_2$ abundance. In Sect. \ref{sec_results} we present the results of the dynamical models, first for a standard metallicity and then as a function of the metallicity, and propose a new formula for the total atomic hydrogen column density. In Sect. \ref{sec_discussion} we discuss our semi-analytical results and compare them to observations and to Hydra PDR Code models.

    \section{\label{sec_analytic_overview}Analytic overview}
        In this section, we present an analytic model including the most important processes controlling the H/H$_2$ transition in interstellar clouds exposed to FUV radiation fields and considering the propagation of the ionization front.
        \subsection{\label{sec_model_geometry}Model geometry}
            We consider a 1D plane-parallel semi-infinite slab of neutral gas irradiated by stellar UV photons. The ionization front, from where we start our model, is assumed punctual similarly to other analytical models: \citep{Sternberg14,Krumholz08,Krumholz09} or to usual PDR codes: Meudon PDR Code, \citep{LePetit2006}, so that the EUV component, responsible for the ionization of hydrogen up to the ionization front, is assumed to be fully extinguished, and the radiation field penetrating the slab is limited to FUV photons, which will create a photodissociation front of H$_2$.
            
            The photoevaporation dynamics of the ionized gas allows the ionization front to advance into the neutral gas with a velocity $v_\mathrm{IF}$. In the reference frame of the ionization front, this is equivalent to an advection of the neutral gas through the PDR and the ionization front with an equal velocity in the opposite direction (toward the ionization front). To account for the impact of this dynamic, we include a uniform advection velocity toward the ionization front, in a similar fashion to the study of \cite{Storzer98}. We assume that the shock front preceding the ionization front has already separated enough from the ionization front so that the H/H$_2$ transition is included in the shocked layer. For the purpose of our model describing the H/H$_2$ transition, we can thus assume a constant advection velocity in the neutral region.
            This velocity is taken as a free parameter, and the possible values and mechanisms controlling this velocity are discussed in Sect. \ref{sec_discussion}.
            
            We also assume the neutral region to have a uniform density $n_{\mathrm{H}}$, irradiated by a unidirectional flux of dissociating photons from the $912-1108$ \AA~ Lyman-Werner (LW) band. In this study, we assume stationary state, since the timescale to reach a stationary PDR structure $t_{\mathrm{eq}} \approx 500~\mathrm{yr}~\left[\left( 10^{6}~\mathrm{cm}^{-3}\right)/n_{\mathrm{H}}\right]$ \citep{Hollenbach1995} is short compared to a cloud lifetime. This timescale is derived from a static view, but gets shorter when the propagation of the ionization front is taken into account as seen in Sect. \ref{sec_physical_model}. We can therefore use the stationary state hypothesis.
            
        \subsection{\label{sec_physical_model}Physical model}
            To consider the impact of photoevaporation on the H/H$_2$ transition, it is thus necessary to expand on previous studies that considered the H/H$_2$ transition in a static slab of gas. The propagation of the ionization front takes the form of an advection velocity of the neutral gas. We do not solve the dynamics of the gas through the ionization front, but only account for the impact of this dynamics by assuming a steady flow with a constant advection velocity. We can then solve for the stationary but out-of-equilibrium state of the H/H$_2$ chemistry in the presence of this steady flow.
        
            Aside from advection, the processes taken into account for the formation and destruction of H$_2$ are its formation on dust and photodissociation by FUV photons. Similarly to \cite{Sternberg14}, we neglect here the impact of cosmic rays, which would maintain a very small fraction of atomic hydrogen deep inside the cloud. In the following, we use most of \cite{Sternberg14} notation conventions.
            
            We take the rate of the H$_2$ formation on dust per hydrogen nucleon as the average value for diffuse gas of \cite{Jura74}, that we assume to be linearly proportional to the metallicity:
            \begin{equation}
                R = 3\times 10^{-17} \, Z'~\mathrm{cm}^3 \,\mathrm{s}^{-1},
                \label{eq_H2_formation_on_dust}
            \end{equation}
            where $Z'$ is the metallicity defined as the ratio between heavy elements abundances in the cloud and in the solar photosphere.
        
            Let $N$ be the hydrogen nuclei column density so that $N = N_1 + 2N_2$ with $N_1$ the column density of atomic hydrogen and $N_2$ the column density of H$_2$. We also introduce $n_1$ and $n_2$ the number densities of H and H$_2$ respectively. For the photodissociation of H$_2$, we assume that the dissociating FUV radiation is only absorbed by grains and H$_2$ (we neglect absorption by gas phase species such as atomic hydrogen, carbon or CO, refer to Appendix \ref{Appendix_C_extinction} for more details) 
            so that the H$_2$ photodissociation rate at a column density $N$ is: 
            \begin{equation}
                \mathcal{P}_{\mathrm{H}_{2}}(N) = \mathcal{P}_0\, G_0\, f_{\mathrm{shield}}(N_2)\, e^{-\sigma_gN},
                \label{eq_H2_photodissociation_rate}
            \end{equation}
            with 
            \begin{equation}
                \mathcal{P}_0 = 3.3 \times 10^{-11}~\mathrm{s}^{-1},
                \label{eq_photodissociation_rate_cst}
            \end{equation}
            and
            \begin{equation}
                \sigma_g = 1.9 \times 10^{-21}\, Z' ~\mathrm{cm}^2,
                \label{eq_dust_absorption_cross_section}
            \end{equation}
            where $G_0$ is the scaling factor of the Interstellar Radiation Field (ISRF) in Habing flux unit, so that $G_0=1$ corresponds to a FUV flux $F_{\mathrm{H}} = 1.27 \times 10^{7} ~\mathrm{photons \, cm^{-2} \, s^{-1}}$ at the cloud surface\footnote{$G_0$ is usually defined as the energy density integrated in the $6 - 13.6 ~\mathrm{eV}$ band divided by the Habing unit, which is the integral of the ISRF spectrum \citep{Habing68} and equals $5.29 \times 10^{-14} ~\mathrm{erg} \, \mathrm{cm}^{-3}$ \citep{Draine2011}, this value is used to deduce the flux unit $F_{\mathrm{H}}$. Defined like that, $G_0$ does not represent the same quantity as the one intervening in Eq. (\ref{eq_H2_photodissociation_rate}), because in this case it should be integrated only in the Lyman-Werner band between $11.3 - 13.6 ~\mathrm{eV}$. We still make the hypothesis that they represent the same quantity. Moreover, and also described in \cite{Draine2011}, other estimates of the ISRF were given by \cite{Draine78} giving 1.69 Habing unit and by \cite{Mathis1983} giving 1.14 Habing unit.} (i.e., at a point in a half-free-space that sees the incoming radiation field from $2\pi~\mathrm{sr}$) and $f_{\mathrm{shield}}$ is the H$_2$ self-shielding function. The parameter $\mathcal{P}_0$ is only an approximate value of the $\mathrm{H}_2$ dissociation rate at the cloud surface for $G_0=1$ as the true value depends on H$_2$ ro-vibrational level populations. The value computed by \cite{Sternberg14} with the isotropic Draine ISRF \citep{Draine78} is $\mathcal{P}_0 = 2.9\times 10^{-11}~\mathrm{s}^{-1}$. Our $\mathcal{P}_0$ value is slightly higher because we consider a contribution of the backscattering of interstellar grains. Moreover, $\sigma_g$ defined in Eq. (\ref{eq_dust_absorption_cross_section}) is the dust effective LW-photon absorption cross section per hydrogen nucleus. We take the same value as \cite{Sternberg14}, derived from a standard interstellar extinction curve with a total-to-selective extinction ratio $R_{\mathrm{V}} \equiv A_{\mathrm{V}}/E(B-V) = 3.1$. We take a linear dependence of $\sigma_g$ to metallicity following the latest prescriptions seen in \cite{Draine2007} and \cite{Wiseman2017} and first given in \cite{Franco1986} with a value at solar abundances very close to the one used here. \cite{Sternberg14} also used $\sigma_g$ proportional to metallicity. A constant dust-to-metal ratio for a wide range of metallicities supports this linear dependence, as found by \cite{Zafar2013}, and more recently by \cite{Chiang2021} for nearby galaxies with a ratio of about 0.5. \cite{Mattsson2014} indicate that the dust-to-metal ratio is kept constant by stellar dust production and dust growth, counteracting dust destruction, except perhaps for low metallicities. Finally, our expression for the H$_2$ self-shielding function is taken from \cite{Draine96}: 
            \begin{multline}
                f_{\mathrm{shield}}(N_2) = \dfrac{0.965}{(1+y/b_5)^2} \\+ \dfrac{0.035}{(1+y)^{0.5}}
                \times \exp{\left[-8.5\times 10^{-4} \, (1+y)^{0.5} \right]},
                \label{eq_37_H2_self_shielding}
            \end{multline}
            with $y = N_2 / 5 \times 10^{14}~ \mathrm{cm}^{-2}$, $b_5 = b / 1~ \mathrm{km\,s}^{-1}$, and $b = 2~ \mathrm{km\,s}^{-1}$ being the Doppler broadening parameter.
            
            The expression of the equation of the density of H$_2$ is:
            \begin{equation}
                \dfrac{dn_2}{dt} = R \, n_{\mathrm{H}} \, n_1 - G_0 \, \mathcal{P}_0 \, e^{-\sigma_gN} \, f_{\mathrm{shield}}(N_2) \, n_2 + v_{\mathrm{IF}} \, \dfrac{dn_2}{dx}.
                \label{eq_dynamical_basic_ODE}
            \end{equation}
            where $v_{\mathrm{IF}}>0$ is the velocity of the cloud in the ionization front frame and $n_{\mathrm{H}}$ is the uniform and stationary density of the gas. On the right-hand side, the first term represents H$_2$ formation on dust, the second term is its photodissociation by FUV, taking into account H$_2$ self-shielding and dust absorption, and the third term is the advection in our 1D geometry. The advection is a positive term in this equation as it acts as a source term for H$_2$. This can be interpreted as a stationary flow of H$_2$ coming from the infinite reservoir that is the molecular region.
            
            The timescale to reach a stationary state is inversely proportional to formation, so that the usual static approach ($v_{\mathrm{IF}}=0$) gives $t_{\mathrm{eq}} = 1/(2Rn_{\mathrm{H}}) \approx 500~\mathrm{yr}~\left[\left( 10^{6}~\mathrm{cm}^{-3}\right)/n_{\mathrm{H}}\right]$ as discussed earlier in Sect. \ref{sec_model_geometry}. But in our advection case, the advection acts like another formation term, so that $t_\mathrm{eq}$ is reduced. The time to reach a stationary state gets shorter as the velocity of the ionization front increases.
            
            We introduce $\tau_g = \sigma_g N$ the dust optical depth in the cloud and $\tau_{1,2} = \sigma_g N_{1,2}$ the optical depths due to the dust associated with the atomic H and the H$_2$ gas, respectively. We also introduce $x_{1,2} = n_{1,2}/n_{\mathrm{H}}$ the fractional abundances of atomic H and the H$_2$. We can now rewrite Eq. (\ref{eq_dynamical_basic_ODE}) in dimensionless form and consider the stationary state $\left(\partial/\partial t =0 \right)$:
            \begin{equation}
                0 = 1 - \dfrac{d \tau_2}{d \tau_g}\left(2 + \dfrac{\mathcal{P}_0}{R} \, \dfrac{G_0}{n_{\mathrm{H}}} \, f_{\mathrm{shield}}\left(\dfrac{\tau_2}{\sigma_g}\right) \, e^{-\tau_g}\right) + \dfrac{v_{\mathrm{IF}}\,\sigma_g}{R}\dfrac{d^2 \tau_2}{d \tau_g^2}.
                \label{eq_dynamical_ODE}
            \end{equation}
            In Eq. (\ref{eq_dynamical_ODE}), we observe the appearance of two dimensionless quantities representing the physical conditions. The first is $\mathcal{P}_0 \, G_0 / R \, n_{\mathrm{H}}$, which gives the ratio of the H$_2$ formation timescale to the H$_2$ dissociation timescale. This parameter (multiplied by the H$_2$ self-shielding function) matches the $\alpha G$ parameter from \cite{Sternberg88} and \cite{Sternberg14}. As explained in these papers, it determines the cloud FUV optical depth due to the dust associated with the H {\textsc i} gas. We find a new dimensionless parameter $v_{\mathrm{IF}} \, \sigma_g / R$, which corresponds to the ratio of the H$_2$ formation timescale to the timescale necessary to advect a dust column density equivalent to $\tau_g=1$. The physical conditions of the cloud $G_0$ and $n_{\mathrm{H}}$ only appear as the $G_0/n_{\mathrm{H}}$ ratio in the first dimensionless parameter, while $v_\mathrm{IF}$ appears separately in the second dimensionless parameter. We then present our results as functions of $G_0/n_{\mathrm{H}}$ and of $v_\mathrm{IF}$. 
            
            We now explore both the case with a null ionization front velocity, which we call "static case" in the following (Sect. \ref{sec_static_case}), and the case with a nonzero ionization front velocity, which we call "propagating ionization front case" in the following abbreviated as "advection case" (Sect. \ref{sec_dynamical_case}).
        \subsection{\label{sec_static_case}Static case}
            \begin{figure}
                \includegraphics[width=\hsize,height=10cm]{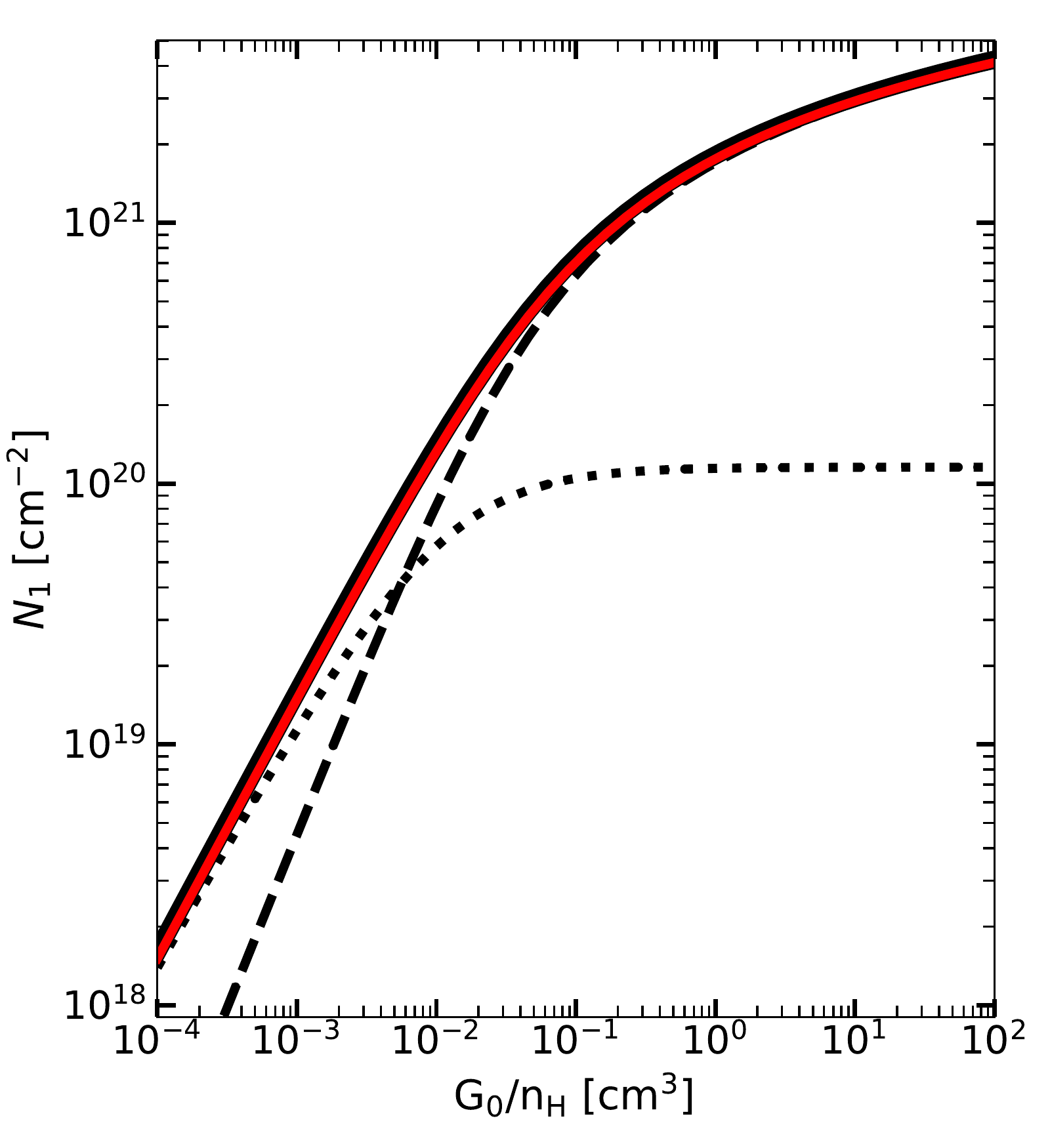}
                \caption{$N_1$ integrated over the atomic region (black dashed line), $N_1$ integrated over the molecular region (black dotted line), $N_{1,\mathrm{tot}}$ (black solid line) as functions of $G_0/n_{\mathrm{H}}$ for a static and stationary approach, calculated by solving Eq. (\ref{eq_static_ODE}) analytically with Eq. (\ref{eq_36_H2_self_shielding}). The red line is $N_{1,\mathrm{tot}}$ from \citeauthor{Sternberg14} \citeyear{Sternberg14} calculated with a similar approach.}
                \label{fig_Analytic_vs_Sternberg}
            \end{figure}
            \begin{figure}
                \includegraphics[width=\hsize]{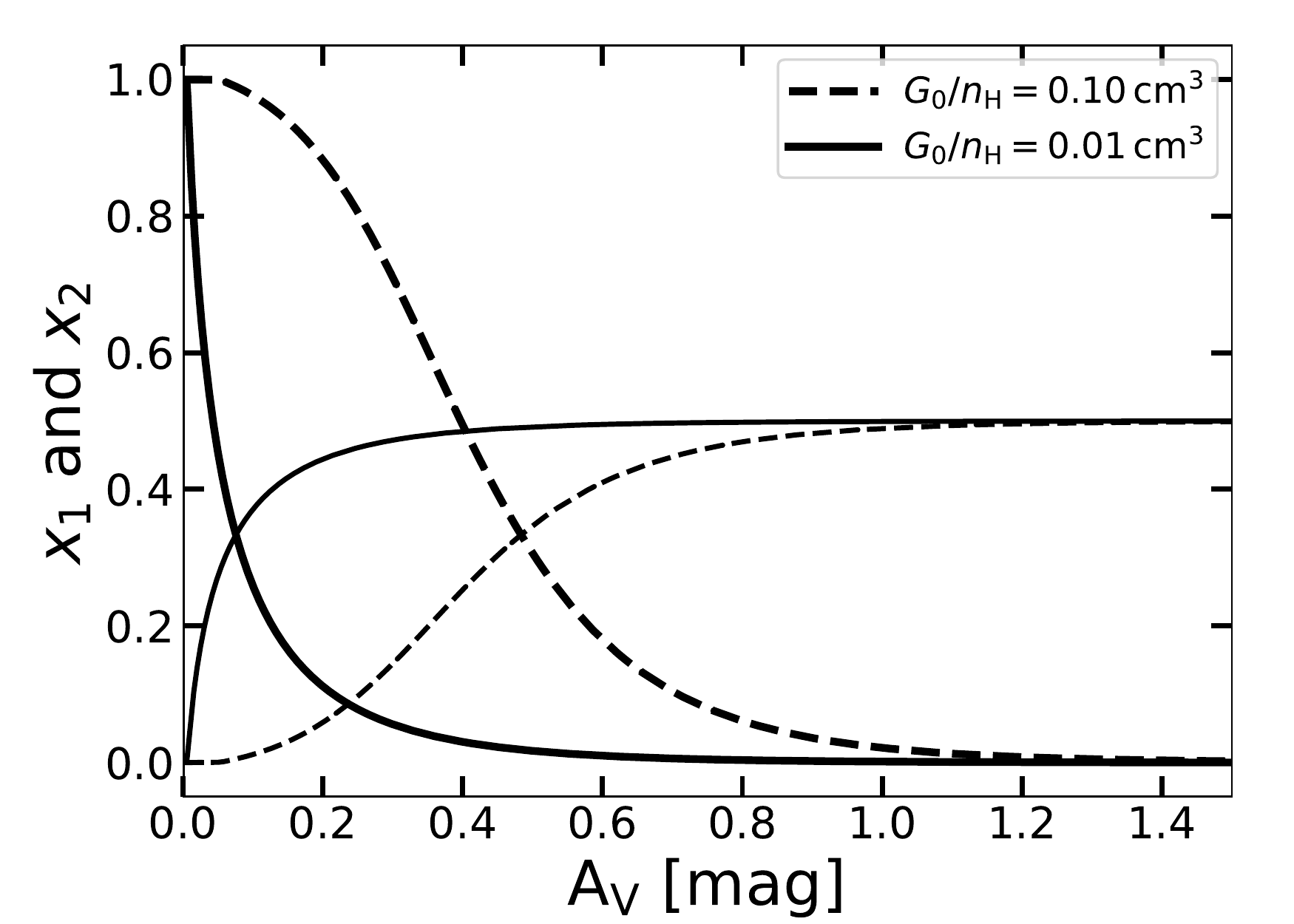}
                \caption{Spatial profiles of H and H$_{2}$ abundances obtained by integrating Eq. (\ref{eq_static_ODE}). The solid 
                lines represent a weak field static solution and the dashed 
                lines represent a strong field static solution. }
                \label{fig_Density_profile_static}
            \end{figure}
            \begin{figure}
                \includegraphics[width=\hsize,height=10cm]{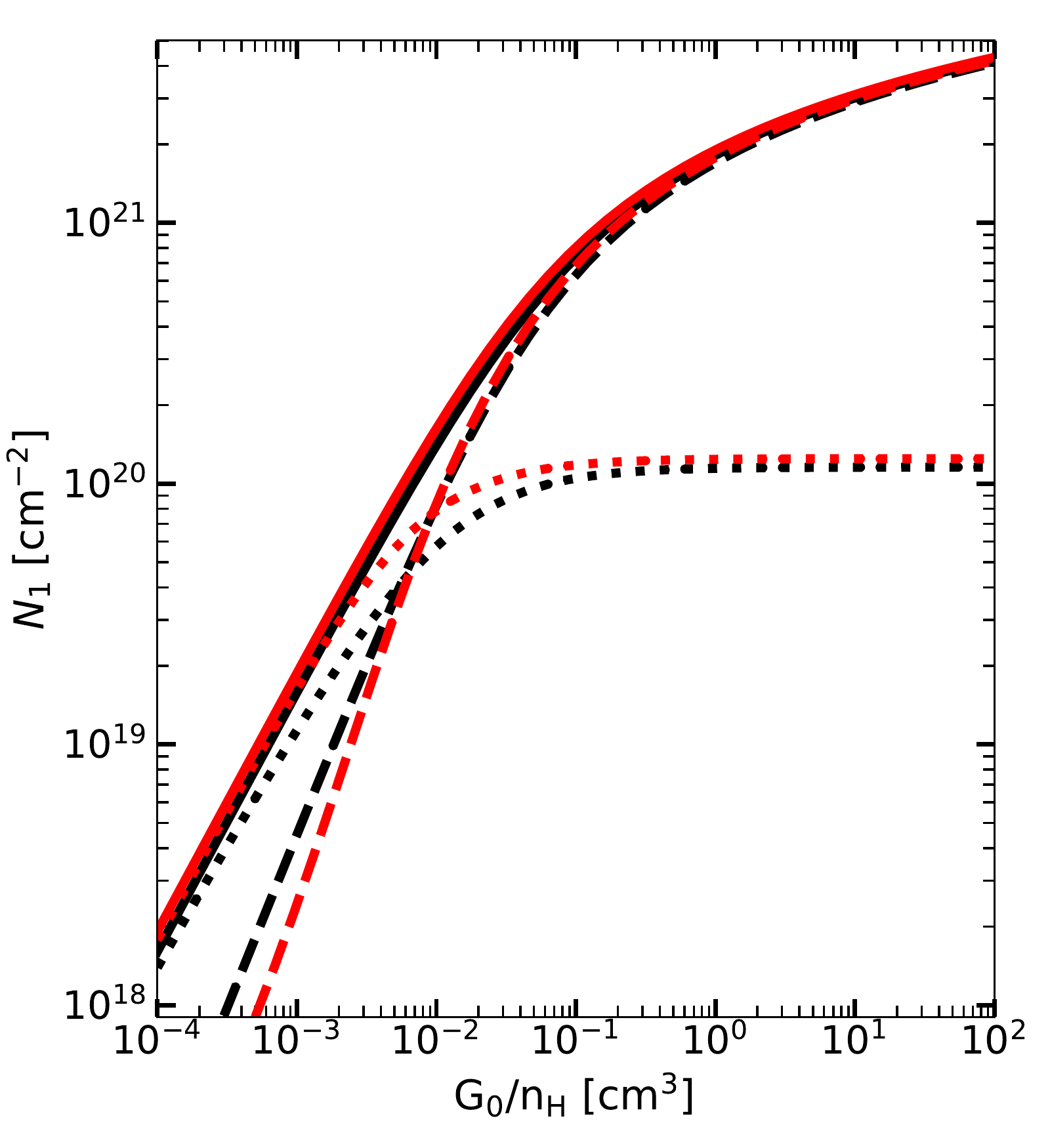}
                \caption{$N_1$ integrated over the atomic region (dashed line), $N_1$ integrated over the molecular region (dotted line), $N_{1,\mathrm{tot}}$ (solid line) as functions of $G_0/n_{\mathrm{H}}$ for a static and stationary approach. Black lines are for Eq. (\ref{eq_36_H2_self_shielding}) of H$_2$ self-shielding, and red lines are for Eq. (\ref{eq_37_H2_self_shielding}), both from \citealt{Draine96}. }
                \label{fig_Analytic_vs_Semi-analytic}
            \end{figure}
            We first rederive the static solution found in \cite{Sternberg14}. In this situation, $v_{\mathrm{IF}} = 0$. Eq. (\ref{eq_dynamical_ODE}) then becomes:
            \begin{equation}
                \dfrac{d \tau_2}{d \tau_g} = \dfrac{1}{2 + \dfrac{\mathcal{P}_0}{R} \, \dfrac{G_0}{n_{\mathrm{H}}} \, \exp{\left(-\tau_g\right)}\,f_{\mathrm{shield}}\left(\tau_2/\sigma_g\right)}.
                \label{eq_static_ODE}
            \end{equation}
            It is possible to solve analytically Eq. (\ref{eq_static_ODE}) by using the following simplified H$_2$ self-shielding function presented in \cite{Draine96}:
            \begin{equation}
                f_{\mathrm{shield}}(N_2) = \left\{ \begin{array}{cl}
                1& \mathrm{~if~} N_2<10^{14}~ \mathrm{cm}^{-2}, \\
                \left( \dfrac{N_2}{10^{14} ~\mathrm{cm}^{-2}} \right)^{-3/4}& \mathrm{~else}.
                \end{array} \right.
                \label{eq_36_H2_self_shielding}
            \end{equation}
            The proof of the solution to Eq. (\ref{eq_static_ODE}) with Eq. (\ref{eq_36_H2_self_shielding}) as the H$_2$ self-shielding function is presented in Appendix \ref{Appendix_Analytic_solution}. We obtained that $\tau_{2}$ can be written as the following function of $\tau_{1}$ by using that $\tau_{g} = \tau_{1} + 2 \tau_{2} $:
            \begin{equation}
                \tau_{2}(\tau_{1}) = \dfrac{1}{2} \Gamma ^{-1} \left[ \dfrac{1}{4},\Gamma \left( \dfrac{1}{4}, \sigma_g \times \left(10^{14}~\mathrm{cm}^{-2} \right)\right) + (1- e^{\tau_{1}})/\xi \right],
                \label{eq_tau2_tau1_static}
            \end{equation}
            with $\Gamma$ the incomplete Gamma function and $\xi$ a dimensionless parameter defined as:
            
            \begin{equation}
                \xi = \dfrac{\left(\sigma_g \times 10^{14}~\mathrm{cm}^{-2} \right)^{3/4}}{\sqrt[4]{2}} \dfrac{\mathcal{P}_0}{R} \dfrac{G_0}{n_{\mathrm{H}}}.
            \end{equation}
            By rearranging Eq. (\ref{eq_tau2_tau1_static}) to have $\tau_{1}\left(\tau_{2}\right)$ and taking the limit $\tau_\mathrm{g}\rightarrow +\infty$ (in which case $\tau_\mathrm{2}\rightarrow +\infty$), we can then obtain the expression for the total H\,{\textsc i} column density $N_{1,\mathrm{tot}}$, also presented in Appendix \ref{Appendix_Analytic_solution}. It gives us :
            \begin{equation}
                N_{1,\mathrm{tot, static}} = \dfrac{1}{\sigma_g}\ln{\left[1 + 2.97 \,  \dfrac{\mathcal{P}_0}{R}\dfrac{G_0}{n_{\mathrm{H}}}\left(\sigma_g\times10^{14}~\mathrm{cm}^{-2}\right)^{3/4} \right]}.
                \label{eq_N_1_tot_static}
            \end{equation}

            
            To finally obtain $\tau_1$ at the position of the transition we need to solve the system given by Eq. (\ref{eq_1_analyt_syst_transit_pos}) (valid only at the position of the transition) and Eq. (\ref{eq_2_analyt_syst_transit_pos}) (valid everywhere). We solved that system numerically to obtain the results shown in Figure \ref{fig_Analytic_vs_Sternberg}. The black solid line represents the total atomic hydrogen column density $N_{1,\mathrm{tot}}$. We introduce two other quantities, which are the column densities of atomic hydrogen in the atomic and molecular regions. The first one is calculated by integrating the atomic hydrogen density only from the beginning of the cloud (i.e., the ionization front) to the H/H$_2$ transition defined as the position where n$_1$ = n$_2$. The second is taken from the transition to the end of the cloud. The sum of these two quantities gives $N_{1,\mathrm{tot}}$. This separation allows to compare the contribution to $N_{1,\mathrm{tot}}$ of the atomic fraction remaining after the H/H$_{2}$ transition. The black dashed line shows the column density of atomic hydrogen integrated in the atomic region. The black dotted line is the column density of atomic hydrogen integrated over the molecular region. A direct comparison of these three curves shows that for the weak field limit\footnote{We define the weak and strong field limits as done in \cite{Sternberg14}, where $\alpha G < 1$ for the weak field limit, corresponding to roughly $G_0/n_{\mathrm{H}} < 10^{-2} ~ \mathrm{cm}^{3}$. For the strong field limit, $\alpha G > 1$, corresponding to $G_0/n_{\mathrm{H}} > 10^{-2} ~ \mathrm{cm}^{3}$.} with $G_0/n_{\mathrm{H}} < 10^{-2} ~ \mathrm{cm}^{3}$, the column density of atomic hydrogen is mainly built up in the molecular region. For the strong field limit with $G_0/n_{\mathrm{H}} > 10^{-2} ~ \mathrm{cm}^{3}$, $N_{1,\mathrm{tot}}$ is mainly from the atomic region. This feature is very well explained in \cite{Sternberg14} as a consequence of the shape of the H/H$_2$ transition. Indeed, on Fig. \ref{fig_Density_profile_static}, one can find the abundances of H and H$_2$ as functions of $A_{\mathrm{V}}$ for two different $G_0/n_{\mathrm{H}}$ ratios. One can see that in the weak-field regime (low $G_0/n_{\mathrm{H}}$ ratio, black solid lines), the atomic hydrogen is present into the molecular medium in significant proportions. In fact, the integral in both the atomic and molecular regions shows that they roughly contribute to the same amount of atomic hydrogen column density. In this weak field regime, the FUV absorption is dominated by H$_2$ self-shielding. In the strong-field regime in dashed lines, the atomic medium has a far higher contribution to the atomic hydrogen column density than the molecular region. In such conditions, the dominant source of far-UV absorption is dust extinction. 
            
            We also show in Fig. \ref{fig_Analytic_vs_Sternberg} the solution of \cite{Sternberg14}. The small difference between the black line from our study and the red one from \cite{Sternberg14} comes from the value of $\mathcal{P}_0$ used and discussed above.
        
            To test the impact of the second and more accurate expression of the shielding function from \citeauthor{Draine96} (Eq. \ref{eq_37_H2_self_shielding}), we solve numerically Eq. (\ref{eq_static_ODE}) and compare the solutions obtained with the two expressions of the self-shielding function in Fig. \ref{fig_Analytic_vs_Semi-analytic}. We see on Fig. \ref{fig_Analytic_vs_Semi-analytic} that the more accurate function of H$_2$ self-shielding as the red curves leads to slight differences, mainly concerning the distribution of the atomic hydrogen before and after the transition. Apart from a little factor in the weak-field regime, the total atomic hydrogen column density is very similar with the two H$_2$ self-shielding functions. We now only use Eq. (\ref{eq_37_H2_self_shielding}) as the H$_2$ self-shielding function in the following.
        \subsection{\label{sec_dynamical_case} Propagating ionization front case}
            To study the case where the ionization front propagates into the cloud (from now on, $v_{\mathrm{IF}} \neq 0$, referred to as the "advection case" in the following), Eq. (\ref{eq_dynamical_ODE}) is solved numerically. As we saw in Eq. (\ref{eq_dynamical_ODE}) the solution does not depend on $G_0$ and $n_{\mathrm{H}}$ independently but on the $G_0/n_{\mathrm{H}}$ ratio. So, the free parameters that are explored are $G_0/n_{\mathrm{H}}$, $Z'$ (appearing in both $R$ and $\sigma_g$ expressions) and $v_{\mathrm{IF}}$. The resolution of this equation gives the spatial profile of $\tau_2$ from which we derive $\tau_1$, and then $x_1$ and $x_2$.
            
            We take $v_{\mathrm{IF}}$ as a free parameter, but as explained in \citealt{Bertoldi96}, this quantity is directly linked to the flux of ionizing photons and so depends on the stellar type of the illuminating star. We discuss this point more thoroughly in Sect. \ref{sec_link_ratio_v_IF}. The results of our numerical computations are presented in Sect. \ref{sec_results}.
        
    \section{\label{sec_results}Results}
        \subsection{\label{sec_advection_effect}Advection effect on the transition}
            \begin{figure}
                \includegraphics[width=\hsize]{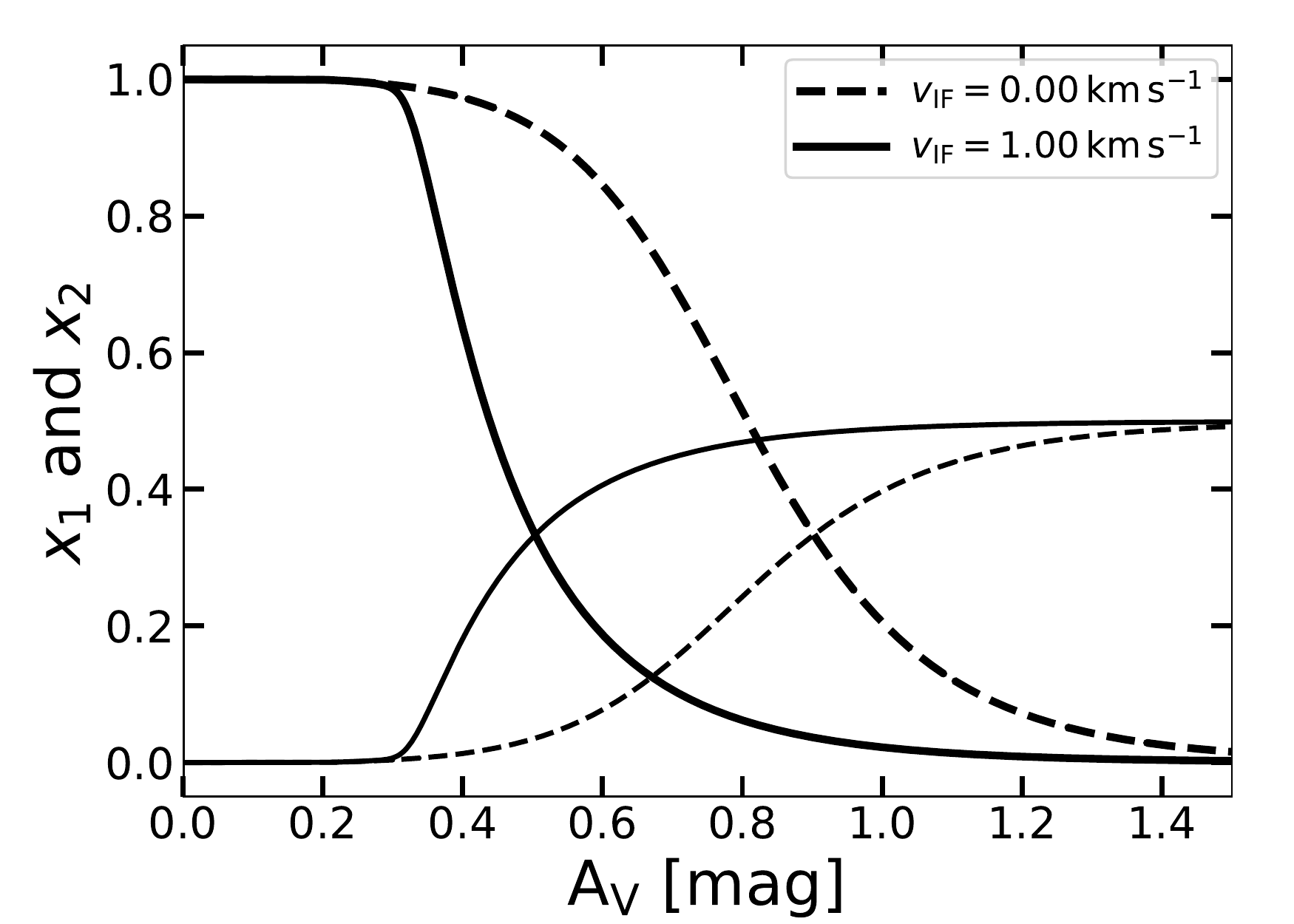}
                \caption{Spatial profiles of H and H$_{2}$ abundances obtained by integration of Eq. (\ref{eq_dynamical_ODE}) with typical PDR conditions, i.e., $G_0/n_{\mathrm{H}} = 0.5 ~ \mathrm{cm}^{3}$ and $v_{\mathrm{IF}} = 1 ~\mathrm{km \,s}^{-1}$. The dashed lines are the static solution and the solid lines are the advection one. }
                \label{fig_Density_profile_dynamical}
            \end{figure}
            The profile of the abundances of H and H$_2$ have been numerically computed by a direct integration of Eq. (\ref{eq_dynamical_ODE}). Until Sect. \ref{sec_metallicity}, we assume solar metallicity $Z' = 1$. We can see an example of the shift between the static and advection profiles discussed above on Fig. \ref{fig_Density_profile_dynamical}. It compares the spatial profiles of atomic hydrogen and molecular hydrogen with and without advection for physical conditions typical of a bright PDR, with a $G_0/n_{\mathrm{H}}$ ratio of $0.5 ~ \mathrm{cm}^{3}$ and ionization front velocity $v_{\mathrm{IF}} = 1 ~\mathrm{km \,s}^{-1}$. In this case, we clearly see that the H/H$_2$ transition is closer to the ionization front on the left than the static model. The higher the ionization front velocity is, the closer the H/H$_2$ transition is to the left edge of the cloud. The atomic region is therefore smaller as well as the accumulated atomic hydrogen column density. For nonzero ionization front velocity, the dissociation front differs from a weaker radiation field because the transition is sharper. To illustrate that, we can compare the solid lines of Fig. \ref{fig_Density_profile_dynamical} representing the H/H$_2$ transition for a nonzero velocity to the dashed line in Fig. \ref{fig_Density_profile_static} representing the H/H$_2$ transition without advection. These two cases have the H/H$_2$ transition roughly at the same position, but in the advection case the transition is much sharper.
            
        \subsection{\label{sec_merging_fronts}Merging fronts}
            \begin{figure}
                \includegraphics[width=\hsize]{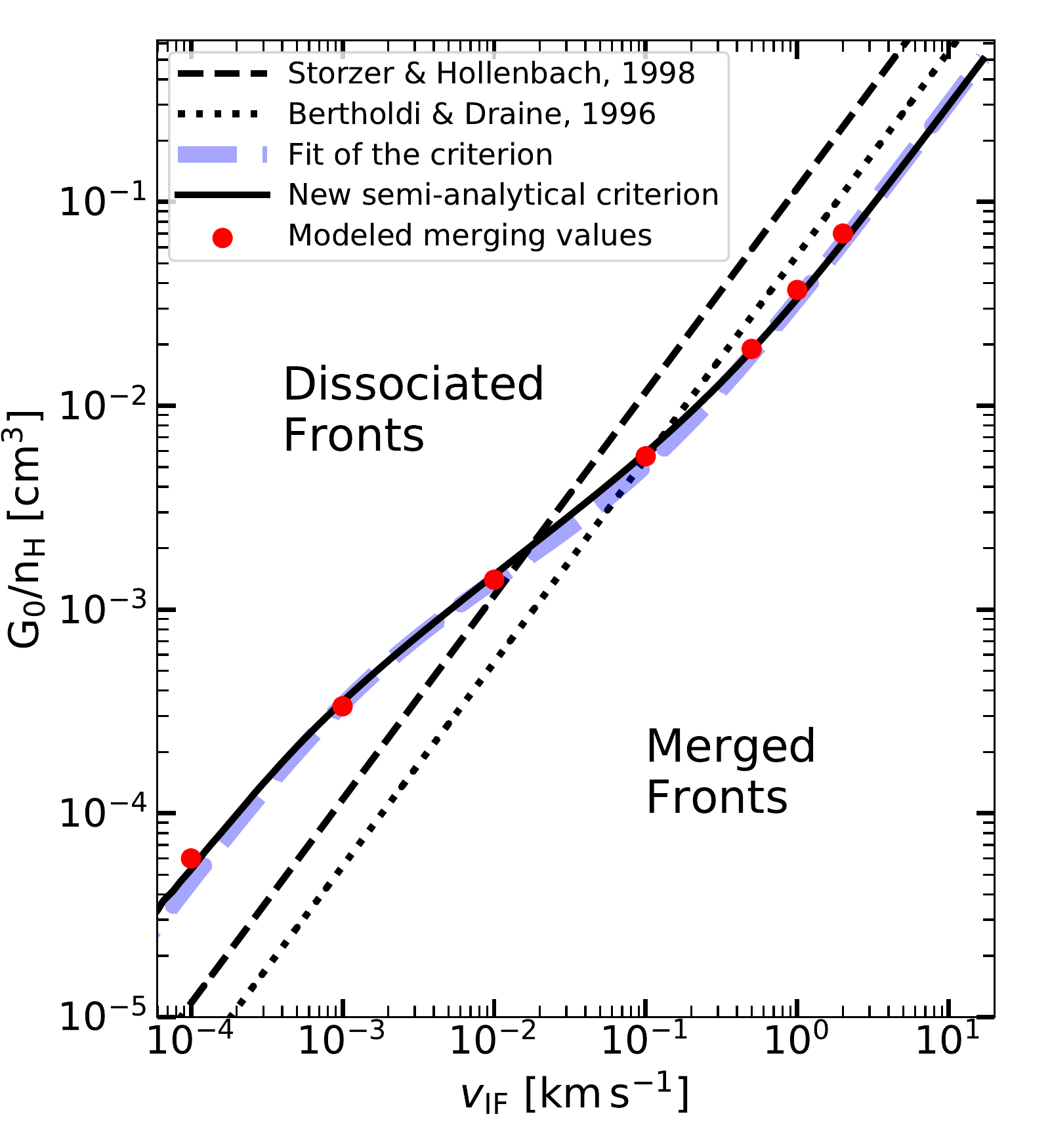}
                \caption{Criterion describing, for a given ionization front velocity, the $G_0/n_{\mathrm{H}}$ ratio leading to a merge of the ionization and dissociation fronts. The ionization front and the dissociation front are merged to the right of the criteria. The black dotted line is the criterion derived by \cite{Bertoldi96}, the black dashed line is the one of \cite{Storzer98}. The red circles are the effective drop from Fig. \ref{fig_N1_static_vs_dynamical}. The black line is our new semi-analytical criterion from Eq. (\ref{eq_merging_criteria}) and the blue dashed line is a fit of the criterion from Eq. (\ref{eq_fit_merging_criterion}).}
                \label{fig_Merging_criteria}
            \end{figure}
            
            With a high-enough velocity, the atomic region disappears and we observe in our models a merging of the dissociation front with the ionization front. This merging was first predicted by \cite{Bertoldi96}. When the ionization front propagates too fast in the medium, the dissociation front cannot separate from the ionization front. In these conditions, they form a merged structure, and the atomic region does not exist. 
        
            We solved Eq. (\ref{eq_dynamical_ODE}) for a grid of different $v_{\mathrm{IF}}$ and different $G_0/n_{\mathrm{H}}$ ratio and determined the critical value of $v_{\mathrm{IF}}$ above which front merging occurs for each value of $G_0/n_{\mathrm{H}}$. These computed critical values are presented as red circles in Fig. \ref{fig_Merging_criteria}. To the right of the boundary defined by these circles, the fronts are merged. In the following we compare the fronts merging criteria from previous papers to our results and derive a new expression for the boundary.
        
            \cite{Bertoldi96} derive their criterion by comparing the velocity of the dissociation front to the velocity of the ionization front assuming they are initially merged. The two fronts can only separate if the initial dissociation front velocity (which depends on $G_0/n_\mathrm{H}$) is larger than the ionization front velocity. In their derivation of the criterion (seen in Fig. \ref{fig_Merging_criteria} as the dotted line) they neglected H$_2$ self-shielding, which is why their criterion is a proportionality relation. 
            
            Another fronts merging criterion was provided in \cite{Storzer98}. To derive their criterion, they compared the dynamical timescale of the flow of H$_2$, representing how long it takes for H$_2$ to flow across the atomic region with the H$_2$ photodissociation timescale. If the dynamical timescale is shorter than the photodissociation timescale, the advected H$_2$ could flow across the PDR without being dissociated and would reach the ionization front, the fronts would be merged. This approach was made without taking into account H$_2$ formation on grains and dust absorption. They also assumed a fixed H$_2$ column density value when taking  H$_2$ self-shielding into account (using Eq. \ref{eq_36_H2_self_shielding}).
            In Fig. \ref{fig_Merging_criteria}, the black dashed line represents \citeauthor{Storzer98} front merging criterion.
            
            As we can see, both of the previously discussed fronts merging criteria are close in term of magnitude, but neither of them can explain the shape of the red circles with the smooth transition between two straight lines, corresponding to two regimes. By not taking into account the H$_2$ self-shielding, \cite{Bertoldi96} are closer to the regime at high values of $v_{\mathrm{IF}}$, whereas \cite{Storzer98}, by taking a fixed value of the H$_2$ self-shielding, are closer to the regime at low-values of $v_{\mathrm{IF}}$. One can guess that the two regimes and the transition between the two come from the treatment of H$_2$ self-shielding. Indeed, for low values of the $G_0/n_{\mathrm{H}}$ ratio, \cite{Sternberg14} explained that H$_2$ self-shielding is the dominant source of FUV absorption, so that advection has a stronger effect in this case. A lower value of $v_{\mathrm{IF}}$ is needed to merge the fronts. We found a closer criterion (the black solid line) by finding an approximation of the slope of $x_2$ at the position of the transition and at the merging of the fronts. We can rewrite Eq. (\ref{eq_dynamical_ODE}) at the position of the transition where $x_1 = x_2 = 1/3$ and when the fronts are merged, namely when $\tau_g = \tau_2 = 0$. With these assumptions, H$_2$ and dust shieldings are equal to $1$, leading to:
            \begin{equation}
                v_{\mathrm{IF}} = \left(\dfrac{\mathcal{P}_0}{R} \dfrac{G_0}{n_{\mathrm{H}}} - 1 \right) \times \dfrac{1}{3} \, \dfrac{R}{\sigma_g} \, \left( \dfrac{dx_2}{d \tau_g}\right)^{-1}.
            \end{equation}
            From here we need to find the $x_2$ derivative by isolating $dx_2/d \tau_g$ in the previous equation and by taking the merging conditions found in the semi-analytical models. The fit gives us:
            \begin{equation}
                \dfrac{dx_2}{d \tau_g} = \dfrac{1}{3} \, \dfrac{1}{4.5} \, \dfrac{\left(\dfrac{\mathcal{P}_0}{R} \dfrac{G_0}{n_{\mathrm{H}}} - 1 \right)}{\exp{\left( \tau_g^0 \right)} - 1},
            \end{equation}
            with $\tau_g^0$ the depth of the H/H$_2$ transition in the static case for this $G_0/n_{\mathrm{H}}$ ratio. Then we can deduce that:
            \begin{equation}
                 v_{\mathrm{IF}} = \dfrac{4.5 \, R}{\sigma_g} \, \left[\exp{\left( \tau_g^0 \right)} - 1\right],
                 \label{eq_merging_criteria}
            \end{equation}
            for which we need to numerically compute $\tau_g^0$ first.
            As it then depends on computing a semi-analytical model, we do not have an analytical expression for the merging criteria. 
            
            To get a more convenient expression for our criterion $\left( G_0/n_{\mathrm{H}} \right)_{\mathrm{crit}}$ as a function of $v_{\mathrm{IF}}$, we fit the merging criterion from Eq. (\ref{eq_merging_criteria}). The fit is presented as the blue dashed line of Fig. \ref{fig_Merging_criteria}, and is given by:
            \begin{multline}
                \left(\dfrac{G_0}{n_{\mathrm{H}}} \right)_{\mathrm{crit}}(v_{\mathrm{IF}}) = 3.0 \cdot 10^{-2} \times \left( \dfrac{v_{\mathrm{IF}} }{1.0~ \mathrm{km \,s}^{-1}}\right) \\
                + 4.5 \cdot 10^{-4} \times \ln{\left(1 + \dfrac{v_{\mathrm{IF}}}{1.0\times 10^{-3} ~ \mathrm{km \,s}^{-1}} \right)}.
                \label{eq_fit_merging_criterion}
            \end{multline}

        \subsection{\label{sec_N1_tot}Total atomic hydrogen column density}
            \begin{figure}
                \includegraphics[width=\hsize]{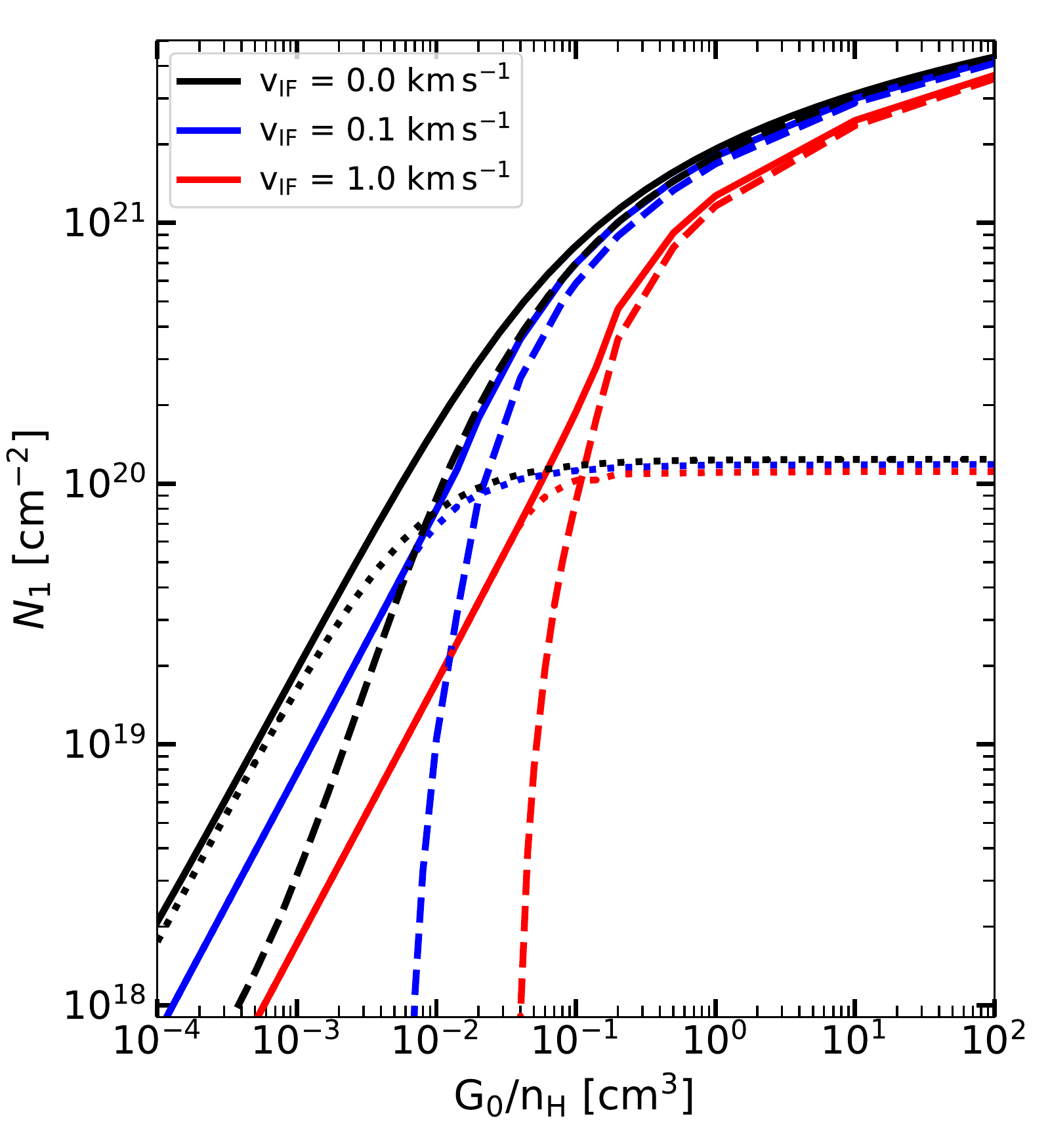}
                \caption{$N_{1}$ integrated over the atomic region (dashed lines), over the molecular region (dotted lines) and over the whole PDR, $N_{1,\mathrm{tot}}$ (solid lines) as functions of the $G_0/n_{\mathrm{H}}$ ratio. The black lines are the static result. The blue and red lines are the advection results respectively for $v_{\mathrm{IF}} = 0.1 ~ \mathrm{and}~ 1.0 ~ \mathrm{km \,s}^{-1}$.}
                \label{fig_N1_static_vs_dynamical}
            \end{figure}
            
            \begin{figure}
                \includegraphics[width=\hsize]{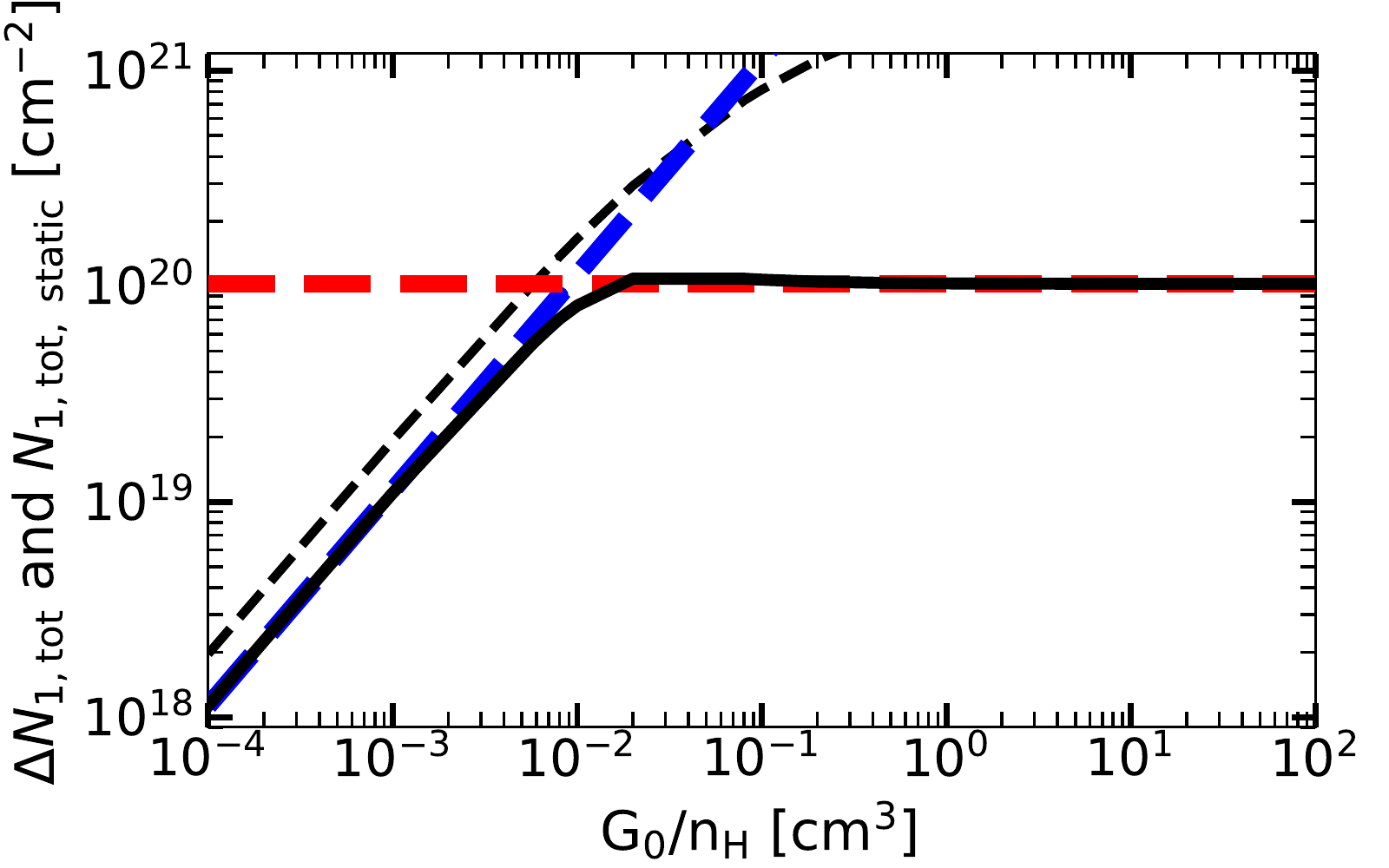}
                \includegraphics[width=\hsize]{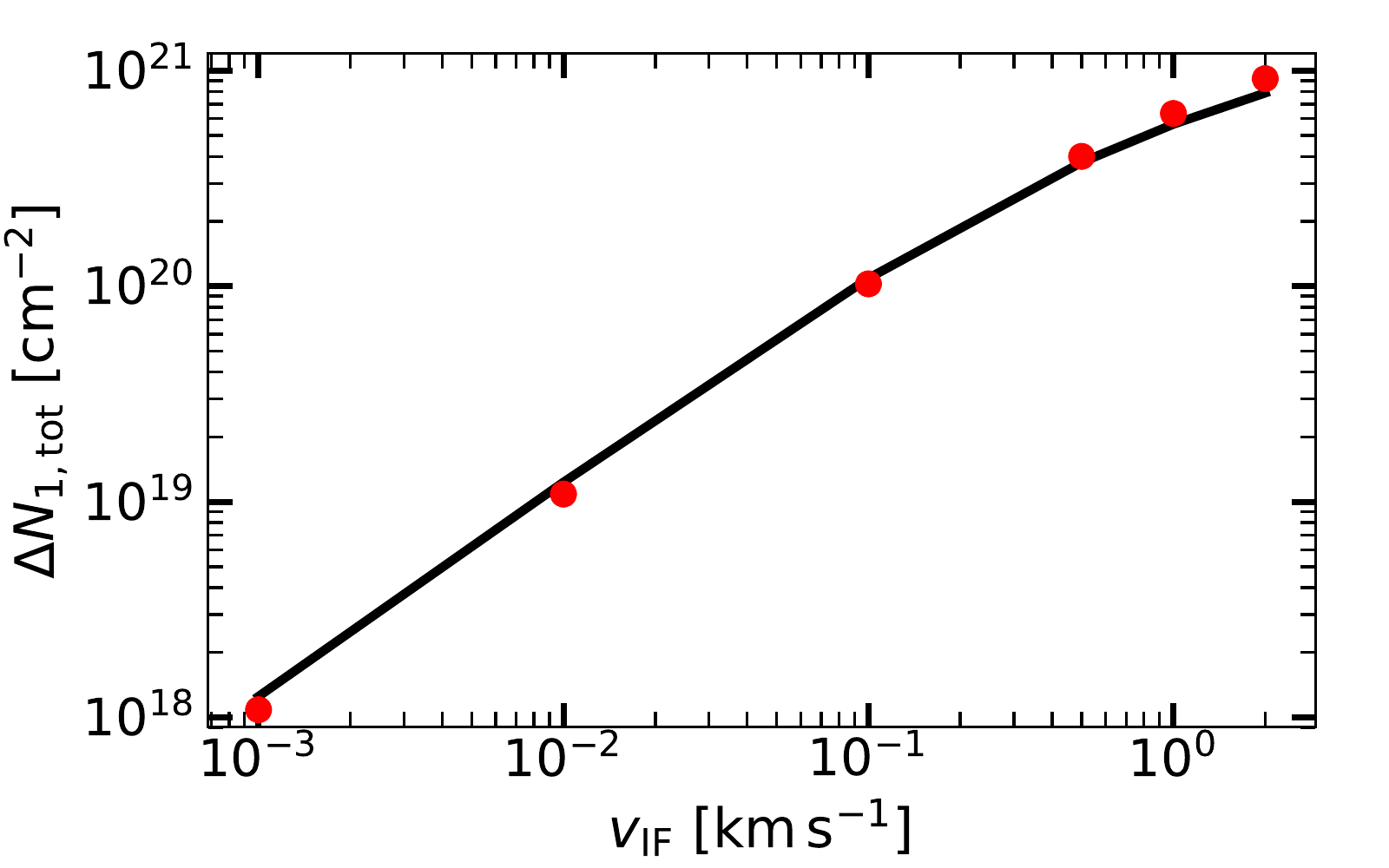}
                \includegraphics[width=\hsize]{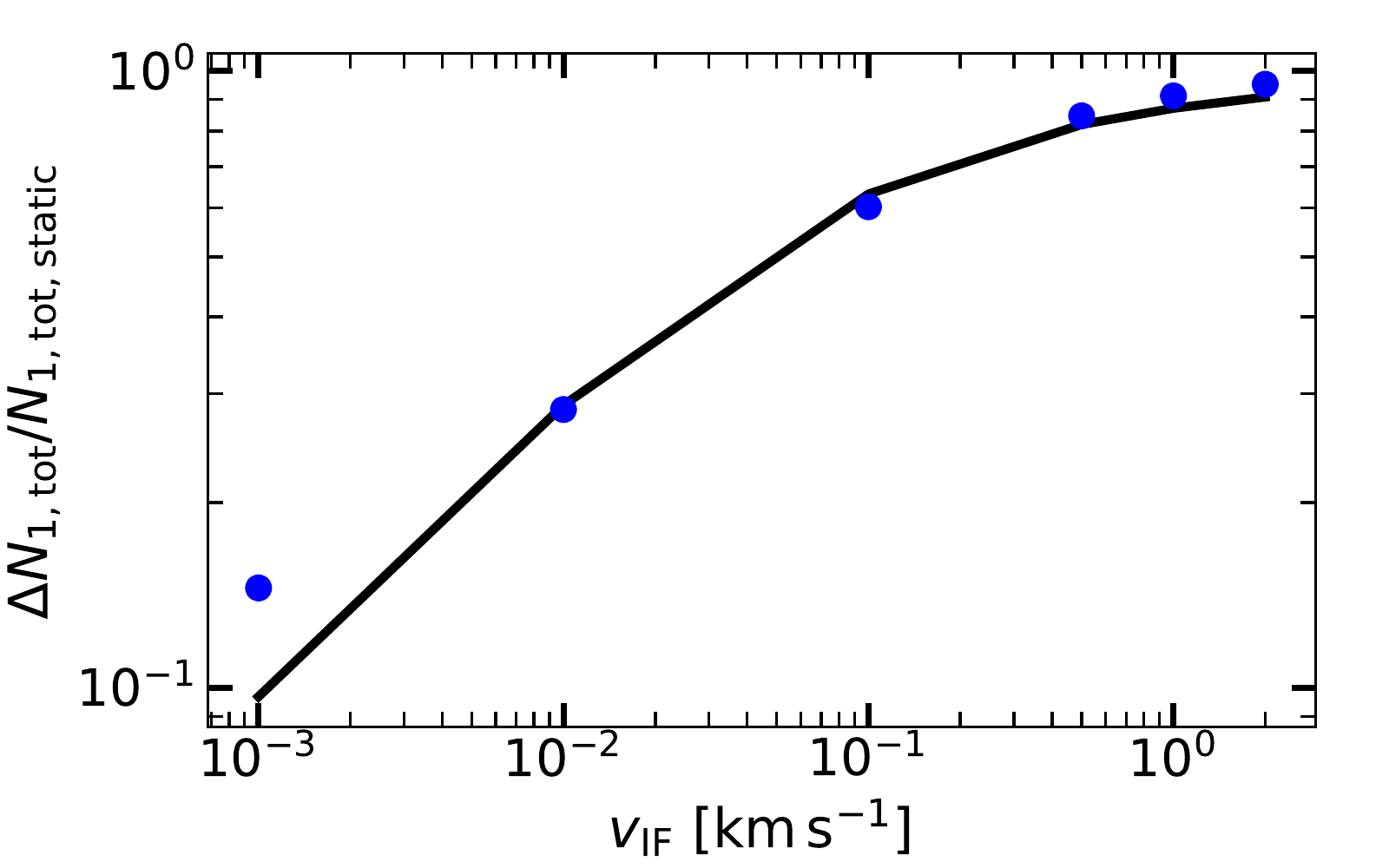}
                \caption{\textit{Top panel:}$\Delta N_{1,\mathrm{tot}}$ (black solid line) as the difference between static (black dashed line for comparison) and advection cases as a function of the $G_0/n_{\mathrm{H}}$ ratio for $v_{\mathrm{IF}}=0.1~\mathrm{km\,s}^{-1}$. The blue and red dashed lines are fits of $\Delta N_{1,\mathrm{tot}}$ in the weak and strong-field limit, respectively. \textit{Middle panel:} $\Delta N_{1,\mathrm{tot}}$ between static and advection cases as a function of $v_{\mathrm{IF}}$. The points were obtained for a strong field of $G_0/n_{\mathrm{H}} = 10^2$. The black lines are the formula presented in Eq. (\ref{eq_shift_N1_tot}). \textit{Bottom panel:} Same as middle panel but for a weak field of $G_0/n_{\mathrm{H}} = 10^{-5}$. In this limit, Eq. (\ref{eq_shift_N1_tot}) is a fit made on all the points, including metallicity, discussed in Sect. \ref{sec_metallicity}.}
                \label{fig_N1_shift}
            \end{figure}
            On Fig. \ref{fig_N1_static_vs_dynamical}, we can see the total atomic hydrogen column density $N_{1,\mathrm{tot}}$ in solid lines, the atomic region contribution to $N_{1,\mathrm{tot}}$ in dashed lines and the molecular region contribution in dotted lines for different velocities ranging from $v_{\mathrm{IF}}=0$ to $1~\mathrm{km \,s}^{-1}$ as functions of the $G_0/n_{\mathrm{H}}$ ratio. We can make a few observations. First and as expected, shifting the transition even slightly toward the ionization front makes $N_{1,\mathrm{tot}}$ decrease from the static case. A smaller shift from a smaller velocity leads to a fainter deviation from the static case. Secondly, we see that for the advection cases, as the $G_0/n_{\mathrm{H}}$ ratio decreases, the atomic region contribution to $N_{1,\mathrm{tot}}$ (in dashed lines) decreases, but the drop is much more abrupt in the advection cases than in the static one. This is the translation of the merging of the fronts discussed above (Sect. \ref{sec_merging_fronts}). As the H/H$_2$ transition approaches and merges with the ionization front, $N_{1}$ integrated in the atomic region drops to $0$. For $G_0/n_{\mathrm{H}}$ ratio under the cutoff, the atomic region does not exist. However, $N_{1,\mathrm{tot}}$ is not 0, because a fraction of atomic hydrogen can still be found in the molecular region, although not being the dominant specie.
        
            We can notice that we retrieve the two regimes discussed in Sect. \ref{sec_static_case} coming from the competition of the two components of $N_{1,\mathrm{tot}}$. The first one is for high $G_0/n_{\mathrm{H}}$ ratio, where $N_{1,\mathrm{tot}}$ is dominated by the atomic hydrogen accumulated in the atomic region. The other is for low $G_0/n_{\mathrm{H}}$, where $N_{1,\mathrm{tot}}$ converges toward a power law representing the atomic hydrogen integrated in the molecular region. The switch between the two regimes follows the merging criteria discussed earlier, so it highly depends on the velocity. As seen on Fig. \ref{fig_Merging_criteria}, higher velocity translates in a switch happening at higher $G_0/n_{\mathrm{H}}$ ratio.
        
            On the top panel of Fig. \ref{fig_N1_shift}, we see $\Delta N_{1,\mathrm{tot}} = N_{1,\mathrm{tot, static}} - N_{1,\mathrm{tot, dynamical}}$ as the black solid line for $v_{\mathrm{IF}}=0.1~\mathrm{km\,s}^{-1}$. The column density $N_{1,\mathrm{tot, static}}$ is in black dashed line for comparison. As we can see, the two regimes discussed above are clearly visible. The change in the total atomic hydrogen column density $\Delta N_{1,\mathrm{tot}}$ is proportional to the $G_0/n_{\mathrm{H}}$ ratio as long as the fronts are merged, and when they are dissociated, $\Delta N_{1,\mathrm{tot}}$ is constant. We highlight this behavior with the superposition of a constant function (red dashed line) to the dissociated fronts regime and a linear function (blue dashed line) to the merged fronts regime.
            
            On the middle panel of Fig. \ref{fig_N1_shift}, we show as red circles the value of $\Delta N_{1,\mathrm{tot}}$ in the dissociated fronts regime as a function of the ionization front velocity $v_{\mathrm{IF}}$. The models were computed for a very high $G_0/n_{\mathrm{H}}$ ratio ($100\,\, \textrm{cm}^3$). The black line is a formula only depending on $v_{\mathrm{IF}}$. On the bottom panel of Fig. \ref{fig_N1_shift}, the blue circles give the $\Delta N_{1,\mathrm{tot}} / N_{1,\mathrm{tot},\mathrm{static}}$ ratio in the merged fronts regime (where $\Delta N_{1,\mathrm{tot}} / N_{1,\mathrm{tot},\mathrm{static}}$ is roughly constant) as a function of $v_\mathrm{IF}$. Here these values were computed with a very low $G_0/n_{\mathrm{H}}$ ($10^{-5}\, \textrm{cm}^3$). The black line is a fit made on the blue circles as well as low-metallicity results as introduced and explained later in Sect. \ref{sec_metallicity}. Looking at Fig. \ref{fig_Merging_criteria}, we see that for such a low $G_0/n_{\mathrm{H}}$ ratio, every velocities investigated lead to merged fronts. But if $v_{\mathrm{IF}}$ is taken small enough, the fronts start to dissociate so the fit does not apply in that case. That is what we start to see with the point where $v_{\mathrm{IF}} = 10^{-3}~\mathrm{km\,s}^{-1}$: our fit does not account very well for very low velocities because the fit is built to replicate strongly merged fronts conditions. Finally, we present the value of $\Delta N_{1,\mathrm{tot}}$ in each of the two regimes (separated fronts and merged fronts) in the following Eq. (\ref{eq_shift_N1_tot}):
        
            \begin{multline}
                \Delta N_{1,\mathrm{tot}} \left[\mathrm{cm}^{-2} \right] = N_{1,\mathrm{tot,static}} - N_{1,\mathrm{tot, dynamical}} = \\
                \left\{ \begin{array}{l}
                    \dfrac{1}{1.5\,\sigma_g}\ln{\left(1 + \dfrac{ v_{\mathrm{IF}} \,\sigma_{\mathrm{g}}}{2\,R}\right)} ~\mathrm{for}~\dfrac{G_{0}}{n_{\mathrm{H}}}>\left(\dfrac{G_0}{n_{\mathrm{H}}} \right)_{\mathrm{crit}}\left( v_{\mathrm{IF}}\right), \\
                    \\
                    \dfrac{N_{1,\mathrm{tot,static}}}{\pi/2} \, \arctan{\left(\sqrt{\dfrac{v_{\mathrm{IF}}}{0.05~\mathrm{km}\, \mathrm{s}^{-1}}}\right)} ~ \mathrm{for}~\dfrac{G_{0}}{n_{\mathrm{H}}}<\left(\dfrac{G_0}{n_{\mathrm{H}}} \right)_{\mathrm{crit}}\left( v_{\mathrm{IF}}\right).
                \end{array} \right.
                \label{eq_shift_N1_tot}
            \end{multline}
            
        \subsection{\label{sec_metallicity}Metallicity}
            \begin{figure}
                \includegraphics[width=\hsize]{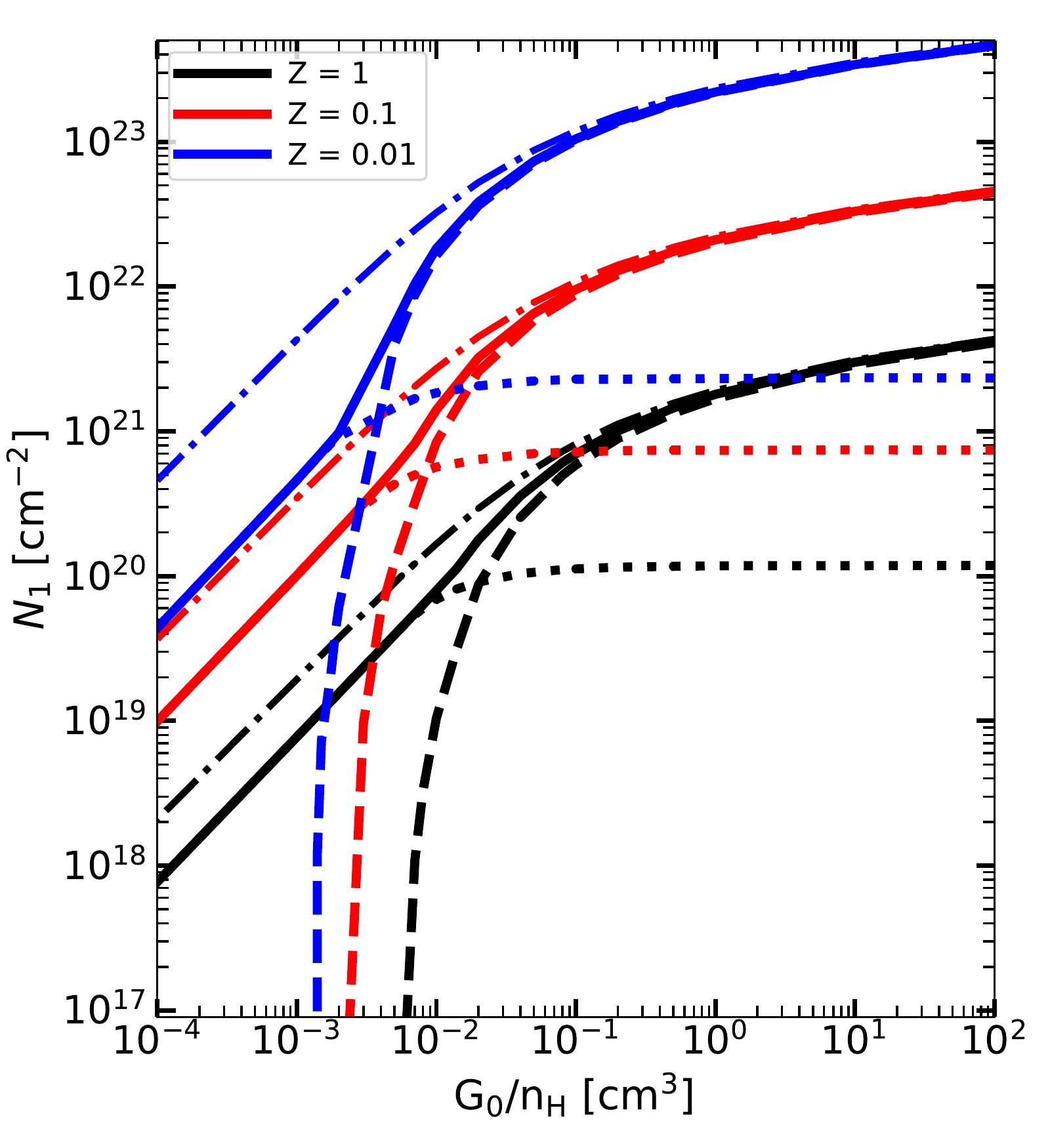}
                \caption{$N_{1}$ integrated over the atomic region (dashed lines), over the molecular region (dotted lines) and over the whole PDR, $N_{1,\mathrm{tot}}$ (solid lines) as functions of the $G_0/n_{\mathrm{H}}$ ratio for, from top to bottom, respectively, $Z' = 0.01$ in blue, $0.1$ in red and $1$ in black, obtained for $v_{\mathrm{IF}} = 0.1~ \mathrm{km \,s}^{-1}$. The dotted-dashed lines are the static results.}
                \label{fig_N1_static_vs_dynamical_vs_metallicity}
            \end{figure}
            
            \begin{figure}
                \includegraphics[width=\hsize]{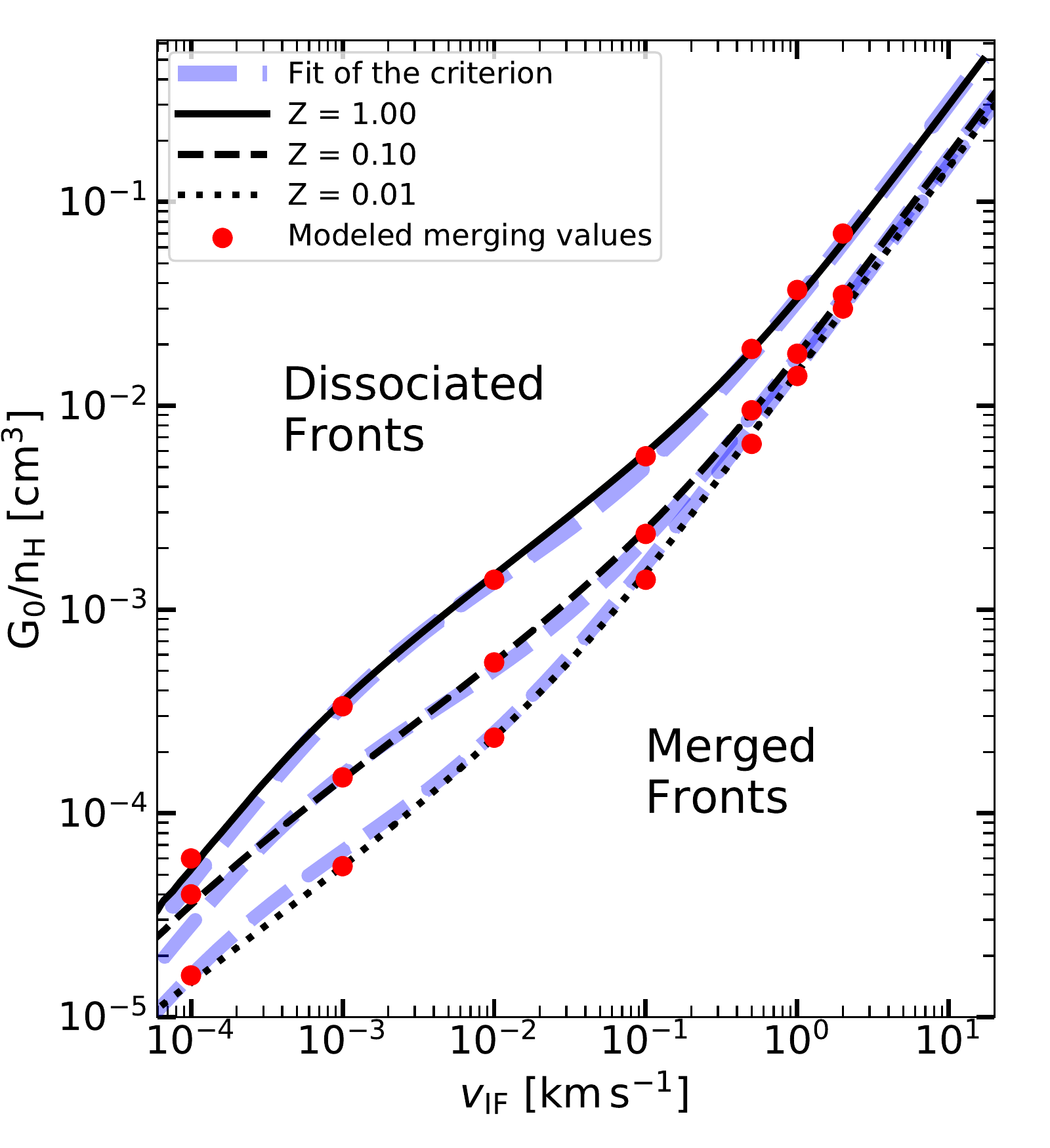}
                \caption{Criterion describing, for a given ionization front velocity, the $G_0/n_{\mathrm{H}}$ ratio leading to a merge of the ionization and dissociation fronts. The red circles are the empirical merging values. The black lines are our new semi-analytical criterion from Eq. (\ref{eq_merging_criteria}) and the blue dashed lines are the fits from Eq. (\ref{eq_fit_merging_criterion_metallicity}). The ionization front and the dissociation front are merged to the right of the criteria.}
                \label{fig_Merging_criteria_metallicities}
            \end{figure}
            
            From Eqs. (\ref{eq_H2_formation_on_dust}) and (\ref{eq_dust_absorption_cross_section}), we can see that by decreasing the metallicity $Z'$, $R$ and $\sigma_g$ are also decreased because of their proportionality to $Z'$. When both of these equations are injected in Eq. (\ref{eq_dynamical_ODE}), we first see that a smaller $R$ increases our dimensionless parameter $\mathcal{P}_0 G_0/ R n_{\mathrm{H}}$, it is then equivalent to a larger $G_0/n_{\mathrm{H}}$ ratio by a factor $1/Z'$. Metallicity does not affect the advection term because the two $Z'$ dependencies cancel each other in the $R/\sigma_g$ ratio. Because the $\mathcal{P}_0 G_0/ R n_{\mathrm{H}}$ ratio depends on $Z'$ but not the advection, the advection needs to be stronger to merge the fronts, causing the merging criterion from Fig. \ref{fig_Merging_criteria} to be shifted to the right for a lower metallicity.
            
            Fig. \ref{fig_N1_static_vs_dynamical_vs_metallicity}, represents, for $v_{\mathrm{IF}} = 0.1~ \mathrm{km \,s}^{-1}$, the atomic hydrogen column density integrated in the atomic region (dashed lines), the molecular region (dotted lines), and the total (solid lines), along with the static models (dotted-dashed lines), for different metallicities $Z' = 0.01, 0.1, 1$, respectively in blue, red, and black. The column density $N_{1,\mathrm{tot}}$ is larger as $Z'$ decreases. Indeed, the parameter $Z'$ acts like a scaling factor because $\sigma_g$ intervenes in the column density $N_{1}=\sigma_g \tau_{1}$. 
            
            Finally, a decreased metallicity has an impact on H$_2$ self-shielding, as the self-shielding term is $f_{\mathrm{shield}}(\tau_2/\sigma_g)$ and $\sigma_g$ decreases as $Z'$. As a consequence, H$_2$ self-shielding is more important relative to dust-shielding. This effect changes the shape and position of the curvature seen for $Z'=1$ in the merging criterion from Fig. \ref{fig_Merging_criteria}.
            
            We computed the merging criteria for the three metallicities, $Z' = 0.01, 0.1, 1$ and presented them in Fig. \ref{fig_Merging_criteria_metallicities}. The merging values obtained with our semi-analytical model are the red circles. We see that the effect is not only a scaling factor but is a bit more complex, yet is very well reproduced by our analytical formula from Eq. (\ref{eq_merging_criteria}) as the black lines. We can see that merging the fronts requires a stronger ionization front velocity as metallicity decreases. The fit presented in Eq. (\ref{eq_fit_merging_criterion}) does not take into account the metallicity, but can still be used by adding a dependence to metallicity as follows:
        
            \begin{multline}
                \left(\dfrac{G_0}{n_{\mathrm{H}}} \right)_{\mathrm{crit}}(v_{\mathrm{IF}}, \,Z') = 1.5\cdot10^{-2} \,\, \left(1 + Z' \right) \left( \dfrac{v_{\mathrm{IF}} }{1.0~ \mathrm{km \,s}^{-1}}\right) \\
                + 4.5 \cdot 10^{-4} \,\, Z'^{\, 2/3} \,\, \ln{\left(1 + \dfrac{v_{\mathrm{IF}}/\sqrt{Z'}}{1.0\cdot 10^{-3} ~ \mathrm{km \,s}^{-1}} \right)}.
                \label{eq_fit_merging_criterion_metallicity}
            \end{multline}
            
            This fit can be seen in Fig. \ref{fig_Merging_criteria_metallicities} as the blue dashed line.
            
            \begin{figure}
                \includegraphics[width=\hsize]{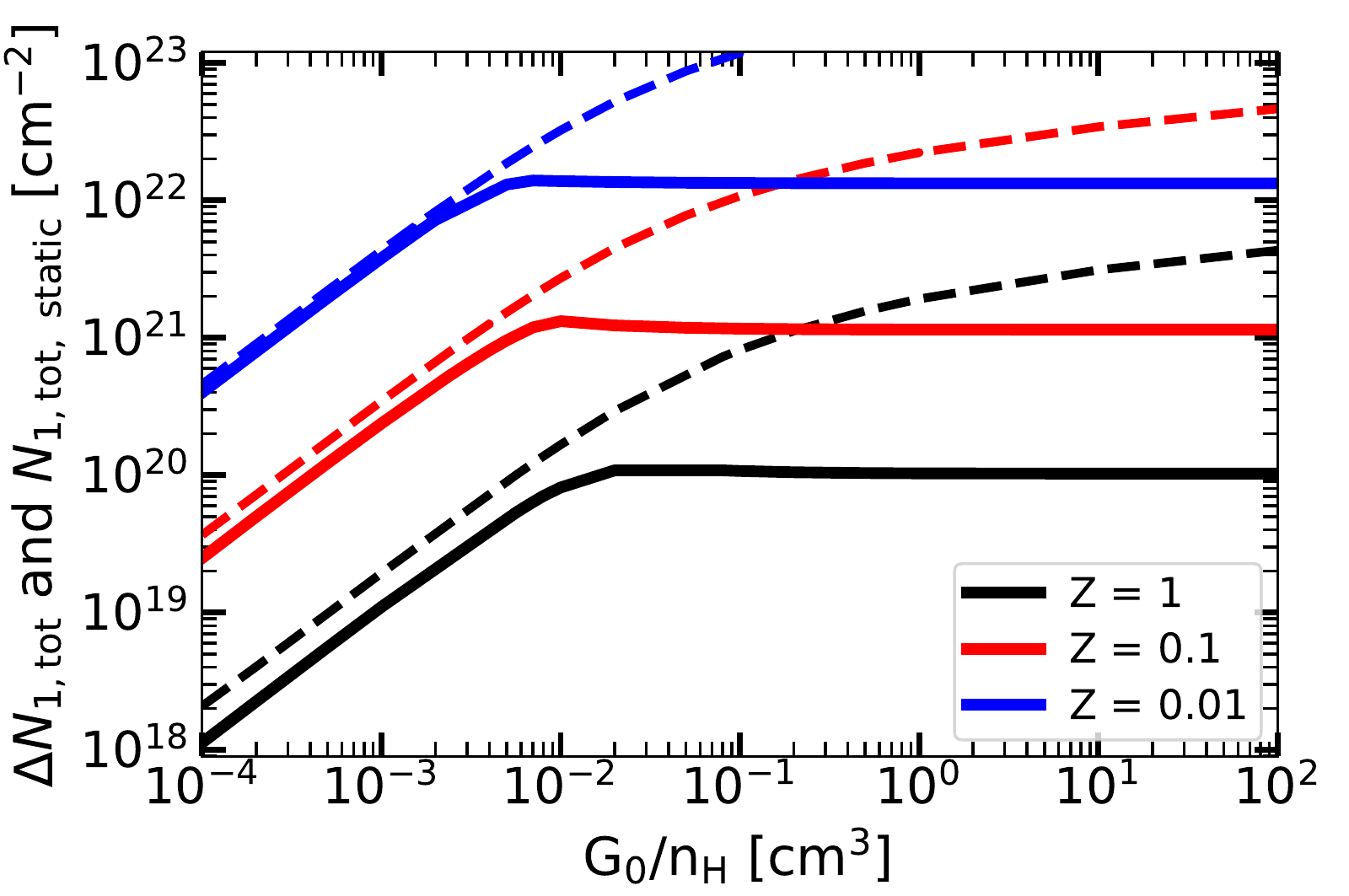}
                \includegraphics[width=\hsize]{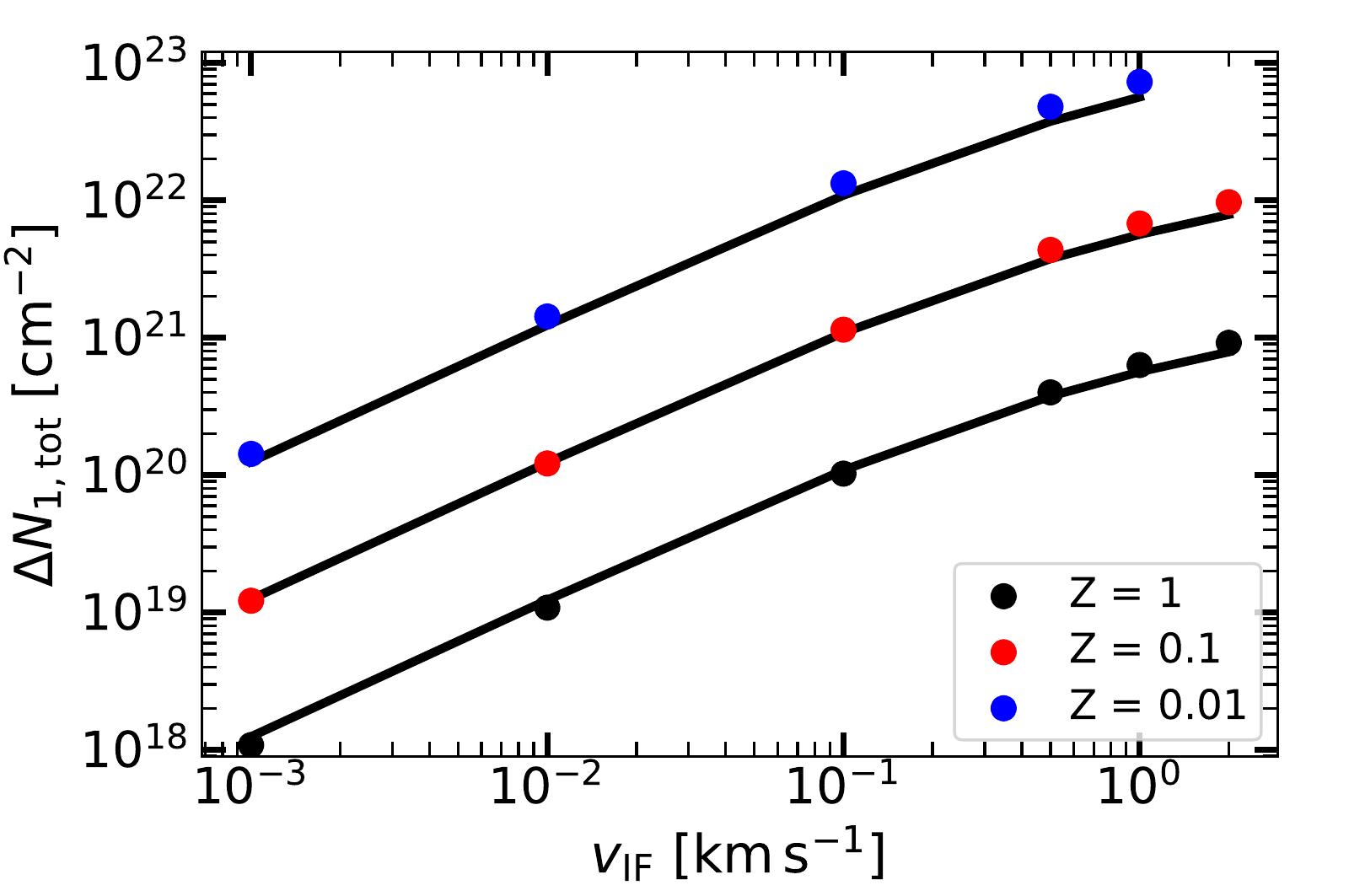}
                \includegraphics[width=\hsize]{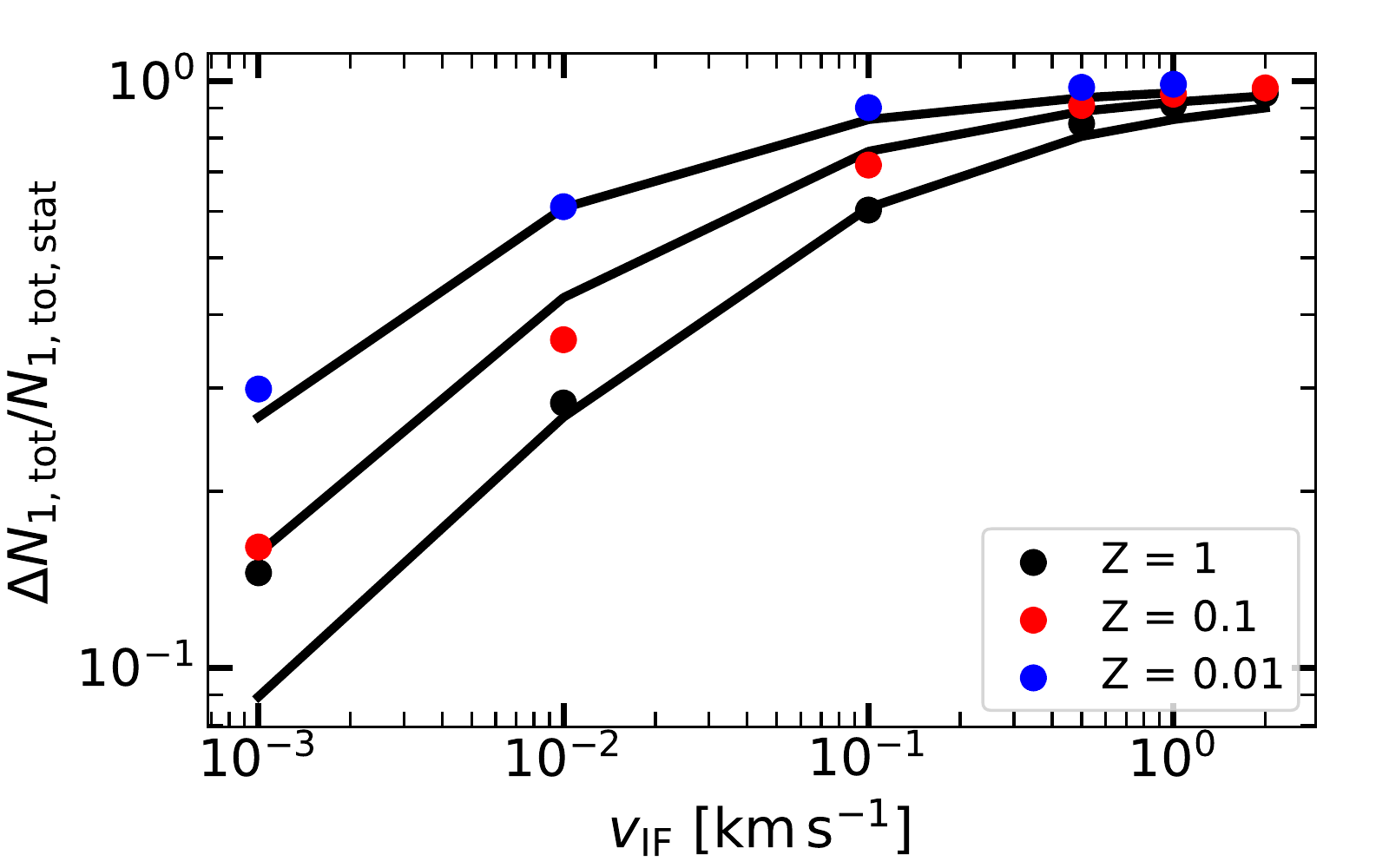}
                \caption{\textit{Top panel:} $\Delta N_{1,\mathrm{tot}}$ (solid lines) as a function of the $G_0/n_{\mathrm{H}}$ ratio for $v_{\mathrm{IF}}=0.1~\mathrm{km\,s}^{-1}$ and for $Z' = 0.01$ (blue), $Z' = 0.1$ (red), and $Z' = 1$ (black). The static $N_{1,\mathrm{tot}}$ are the dashed lines. \textit{Middle panel:} $\Delta N_{1,\mathrm{tot}}$ between static and advection cases as a function of $v_{\mathrm{IF}}$. The points were obtained for a strong field of $G_0/n_{\mathrm{H}} = 10^2\, \textrm{cm}^3$. The black lines are the analytical formula presented in Eq. (\ref{eq_shift_N1_tot_metallicity}). \textit{Bottom panel:} Same as middle panel but for a weak field of $G_0/n_{\mathrm{H}} = 10^{-5}\, \textrm{cm}^3$. In this limit, Eq. (\ref{eq_shift_N1_tot_metallicity}) is a fit made on all the points.}
                \label{fig_N1_shift_metallicity}
            \end{figure}
            
            The two regimes dichotomy discussed previously for the $Z'=1$ case is still applicable to the other cases, as seen on the top panel of Fig. \ref{fig_N1_shift_metallicity}, with the merged or dissociated fronts regimes. We clearly observe the two regimes with a constant value for dissociated fronts and a power law for merged fronts. On the middle panel, we see $\Delta N_{1,\mathrm{tot}}$ for a few metallicities. Our formula from Eq. (\ref{eq_shift_N1_tot}) fits very well the shift in this dissociated fronts regime, reproducing the scaling factor. On the bottom panel, the semi-analytical results were fit to obtain the black curve. We present hereafter the final formula for both of the regimes and including metallicity:
            \begin{multline}
                \Delta N_{1,\mathrm{tot}} \left[\mathrm{cm}^{-2} \right] = N_{1,\mathrm{tot,static}} - N_{1,\mathrm{tot, dynamical}} = \\
                \left\{ \begin{array}{l}
                    \dfrac{1}{1.5\,\sigma_g}\ln{\left(1 + \dfrac{ v_{\mathrm{IF}} \,\sigma_{\mathrm{g}}}{2\,R}\right)} \, \, \scriptstyle{\mathrm{for} \, \, \dfrac{G_{0}}{n_{\mathrm{H}}}>\left(\dfrac{G_0}{n_{\mathrm{H}}} \right)_{\mathrm{crit}}\left( v_{\mathrm{IF}}, Z'\right)}, \\
                    \\
                    \dfrac{N_{1,\mathrm{tot,static}}}{\pi/2} \, \arctan{\left(\sqrt{\dfrac{v_{\mathrm{IF}}/\sqrt{Z'}}{0.05~\mathrm{km}\, \mathrm{s}^{-1}}}\right)} \, \, \scriptstyle{ \mathrm{for}\, \, \dfrac{G_{0}}{n_{\mathrm{H}}}<\left(\dfrac{G_0}{n_{\mathrm{H}}} \right)_{\mathrm{crit}}\left( v_{\mathrm{IF}}, Z'\right)}.
                \end{array} \right.
                \label{eq_shift_N1_tot_metallicity}
            \end{multline}
        
            In the dissociated fronts regime, where $G_{0}/n_{\mathrm{H}}>\left(G_0/n_{\mathrm{H}} \right)_{\mathrm{crit}}\left( v_{\mathrm{IF}}, Z'\right)$ and where $\Delta N_{1,\mathrm{tot}}$ is constant with respect to $G_{0}/n_{\mathrm{H}}$, the effect of metallicity is just a scaling factor that affects our fit through the dependence of $\sigma_g$ on $Z'$. In the merged fronts regime ($G_{0}/n_{\mathrm{H}}<\left(G_0/n_{\mathrm{H}} \right)_{\mathrm{crit}}\left( v_{\mathrm{IF}}, Z'\right)$), the effect or metallicity can be described as a shift to the left of the fit as $Z'$ decreases, modeled by the $\sqrt{Z'}$ factor in the fit.
        
    \section{\label{sec_discussion}Discussion}
        \subsection{\label{sec_link_ratio_v_IF}The link between $G_0/n_{\mathrm{H}}$ and $v_{\mathrm{IF}}$}
            As explained in Sect. \ref{sec_dynamical_case}, we treated $v_{\mathrm{IF}}$ as a free parameter independent of $G_0$. We investigate in this section how $v_{\mathrm{IF}}$ is linked to our $G_0/n_{\mathrm{H}}$ ratio. First, mass conservation across the ionization front (Eq. (37.1), \cite{Draine2011}) yields :
            \begin{equation}
                v_{\mathrm{IF}} = \dfrac{F_{\mathrm{EUV}}(\mathrm{H}^+/\mathrm{H})}{n_{\mathrm{H}}},
                \label{eq_IF_velocity_F_EUV}
            \end{equation}
            with $F_{\mathrm{EUV}}(\mathrm{H}^+/\mathrm{H})$ the flux of ionizing photons arriving at the ionization front. This flux depends on the absorption of EUV photons in the ionized region, on the distance between the star and the PDR and on the spectral type of the star. The $G_0$ parameter similarly depends on the distance of the star, the spectral type of the star and the absorption of FUV photons in the ionized region. 
            
            We note that $G_0/n_{\mathrm{H}}$ and $v_{\mathrm{IF}}$ are determined by taking the EUV and FUV fluxes at the ionization front. The star emission rate of EUV and FUV photons are respectively $S_{\mathrm{EUV}}$ and $S_{\mathrm{FUV}}$ (photons s$^{-1}$) and the distance between the star and the ionization front is $d_{star}$. Following \cite{Bertoldi96}, we call $q_{\mathrm{EUV}}$ the factor representing EUV absorption by dust and hydrogen recombination in the ionized gas, and $q_{\mathrm{FUV}}$ the factor representing FUV dust absorption in the ionized gas. The ionization front velocity is then expressed as \citep{Bertoldi96}:
            \begin{equation}
                v_{\mathrm{IF}} = \dfrac{S_{\mathrm{EUV}}}{4\pi \, d_{star}^{2} \, q_{\mathrm{EUV}} \, n_{\mathrm{H}}},
            \end{equation}
            and the $G_0/n_{\mathrm{H}}$ ratio is:
            \begin{equation}
                \dfrac{G_0 \times F_{\mathrm{ISRF}}}{n_{\mathrm{H}}} = \dfrac{S_{\mathrm{FUV}}}{4\pi \, d_{star}^{2} \, q_{\mathrm{FUV}} \, n_{\mathrm{H}}},
            \end{equation}
            with $F_{\mathrm{ISRF}} = 1.27 \times 10^{7} \mathrm{cm}^{-2} \, \mathrm{s}^{-1}$. Dividing the two expressions and using Eq. (12) from \cite{Bertoldi96}:
            \begin{equation}
                \dfrac{v_{\mathrm{IF}}}{G_0/n_{\mathrm{H}}} = \dfrac{S_{\mathrm{EUV}}}{S_{\mathrm{FUV}}}  \times\dfrac{\exp{\left((\sigma_g-\sigma_{\mathrm{EUV}})N_{\mathrm{HII}}\right)}}{1+ \dfrac{5.07 \times 10^{-20} \times N_{\mathrm{HII}}}{1 + \sigma_{\mathrm{EUV}} \, N_{\mathrm{HII}}/3}} \times F_{\mathrm{ISRF}},
                \label{eq_link_v_IF_ratio}
            \end{equation}
            where $N_{\mathrm{HII}}$ is the total gas column density integrated in the ionized region and $\sigma_{\mathrm{EUV}}$ the EUV dust extinction cross section. In this expression, the link between the ionization front velocity and $G_0/n_{\mathrm{H}}$ is quite clear, depending only on the properties of the star and the absorption by dust and gas before the ionization front. 
            
            As stated in \cite{Bertoldi96}, the result depends a lot on the dust absorption properties. In the following, we use $\sigma_g$ from Eq. (\ref{eq_dust_absorption_cross_section}). For $\sigma_{\mathrm{EUV}}$, the EUV dust extinction, we take the value of \cite{Bertoldi96} for $R_{\mathrm{V}} = 3.1$:
            \begin{equation}
                \sigma_{\mathrm{EUV}} = 1.64 \times 10^{-21}~\mathrm{cm}^2.
            \end{equation}
        
            As we see in Sect. \ref{subsect_observations}, for a various set of objects, $\tau_{\mathrm{EUV}} = \sigma_{\mathrm{EUV}}N_{\mathrm{HII}}$ is close to 1 in a range going from about 0.1 to 10. We decided to take a look at the link between $v_{\mathrm{IF}}$ and the $G_0/n_{\mathrm{H}}$ ratio for this range of EUV dust extinction. As the link also depends on the ratio between the EUV and FUV emission rates, we first assume $S_{\mathrm{EUV}}/S_{\mathrm{FUV}} = 1$, corresponding to an early-type O star.
        
            \begin{figure}
                \includegraphics[width=\hsize]{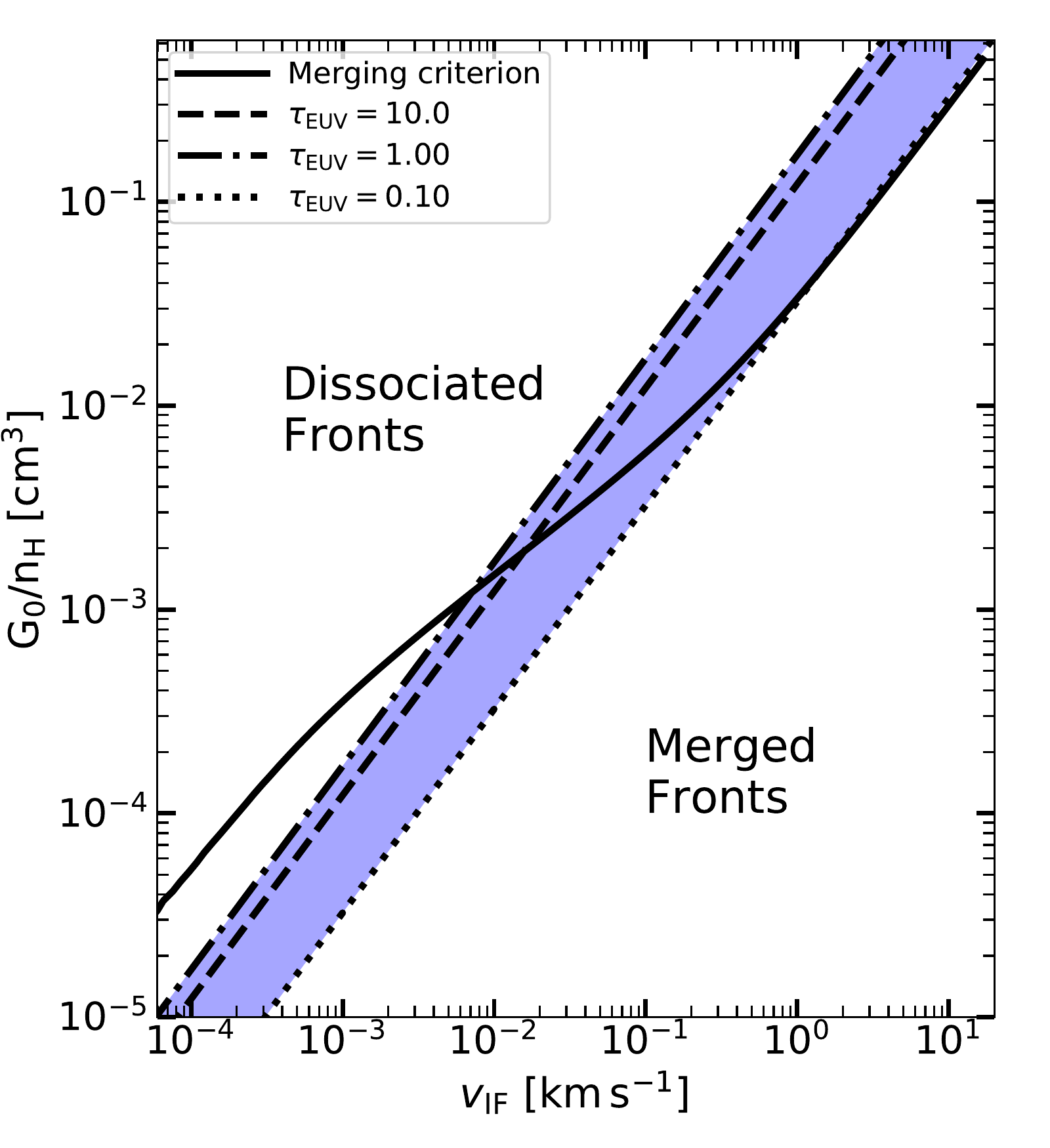}
                \caption{Comparison between the merging criterion from Fig. \ref{fig_Merging_criteria} and the link between $v_{\mathrm{IF}}$ and the $G_0/n_{\mathrm{H}}$ ratio from Eq. (\ref{eq_link_v_IF_ratio}).}
                \label{fig_Merging_criteria_tau_EUV}
            \end{figure}
            
            Fig. \ref{fig_Merging_criteria_tau_EUV} represents the link between $v_{\mathrm{IF}}$ and the $G_0/n_{\mathrm{H}}$ ratio for different $\tau_{\mathrm{EUV}}$, 0.1 in dotted line, 1 in dashed line, and 10 in dotted-dashed line. This range is compared to the merging criterion in solid line. We see that for a low value of $\tau_{\mathrm{EUV}}$ the conditions are always very close to front merging. Indeed, a low $\tau_{\mathrm{EUV}}$ means that EUV absorption is low, resulting in a fast propagation of the ionization front into the cloud. For higher values of $\tau_{\mathrm{EUV}}$, the fronts are merged for weak fields, with $G_0/n_{\mathrm{H}}$ ratio under $10^{-2}~ \mathrm{cm}^{3}$. For stronger fields, the link between $v_{\mathrm{IF}}$ and the $G_0/n_{\mathrm{H}}$ ratio does not move far away from the merging criterion, with less than a decade of difference. In such conditions, fronts are clearly dissociated, yet dynamical effects are still visible on $N_{1,\mathrm{tot}}$.
        
            \begin{figure}
                \includegraphics[width=\hsize]{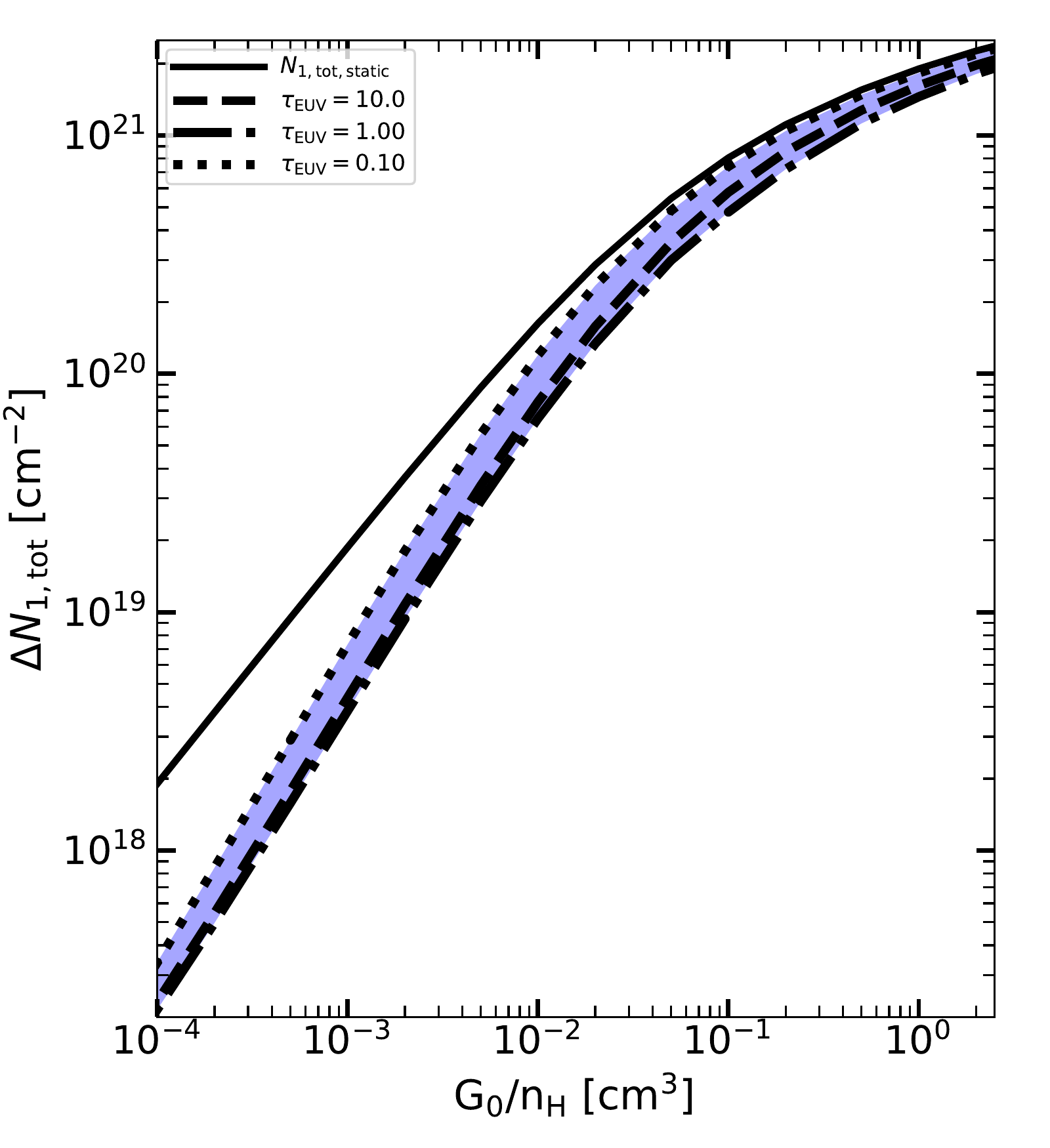}
                \caption{Using the link between $v_{\mathrm{IF}}$ and the $G_0/n_{\mathrm{H}}$ ratio from Eq. (\ref{eq_link_v_IF_ratio}), we present the $\Delta N_{1,\mathrm{tot}}$ range for different $\tau_{\mathrm{EUV}}$ compared with $N_{1,\mathrm{tot}}$ from the static solution. The closer $\Delta N_{1,\mathrm{tot}}$ is to $N_{1,\mathrm{tot,~static}}$, the stronger the dynamical effects. }
                \label{fig_N1_shift_v_IF_vs_ratio}
            \end{figure}
            
            That is what is presented on Fig. \ref{fig_N1_shift_v_IF_vs_ratio} for the same range of $\tau_{\mathrm{EUV}}$. We see that $\Delta N_{1,\mathrm{tot}}$ is very close to the static solution, meaning that dynamical effects are very strong. They are higher for low $\tau_{\mathrm{EUV}}$ values as we are closer to the front merging conditions, but are still very high as $\tau_{\mathrm{EUV}}$ increases. We also observe that the effects are more important in the strong field limit, because for such conditions the ionization front velocity is higher, meaning that the H/H$_{2}$ transition is sharper, translating into a lower atomic fraction in the molecular region.
            
            As we can see in Eq. (\ref{eq_IF_velocity_F_EUV}), the velocity of a steady ionization front is inversely proportional to the gas density immediately behind the IF. In the classical scenario of a D-type front preceded by a shock front, this density is thus not the initial density of the molecular cloud, but the density after shock compression. In addition, density gradients preexisting in the molecular cloud lead either to an accelerating ionization front (propagating in the direction of a decreasing density gradient, as e.g., in the formation of a champagne flow H\,\textsc{ii} region) or a decelerating ionization front (propagating in the direction of an increasing density gradient). As shown by \cite{Franco1989, Franco1990}, respectively for spherically symmetric and disk-like clouds, an ionization front propagating from the inside of the structure is accelerated along the decreasing density ramp, and if the ramp is steeper than $r^{-3/2}$, the D-type ionization front overtakes the shock front and becomes a R-type front. This acceleration originating from the density structure leads to the aforementioned blisters and champagne flows. The stationary hypothesis for the PDR structure only remains valid if the timescale of variation of $n_{\mathrm{H}}$ and $v_{\mathrm{IF}}$ due to the density gradient and the acceleration of the ionization front is larger than the timescale to reach a stationary PDR structure. As $n_{\mathrm{H}}$ and $v_{\mathrm{IF}}$ are linked to each other by Eq. (\ref{eq_IF_velocity_F_EUV}), the variation of one of them causes the variation of the other, so that their timescales of variation are equal. The timescale of variation of the density $n_{\mathrm{H}}$ is:
            \begin{equation}
                \tau_{n_{\mathrm{H}}} = \dfrac{n_{\mathrm{H}}}{dn_{\mathrm{H}}/dt} = \dfrac{n_{\mathrm{H}}}{v_{\mathrm{IF}} \times dn_{\mathrm{H}}/dx}.
            \end{equation}
            By comparing it to the timescale to reach a stationary PDR structure, which is $(2 \, R \, n_{\mathrm{H}})^{-1}$ \citep{Hollenbach1995}, one finds that to maintain a stationary state, the logarithmic derivative of the density has to verify:
            \begin{equation}
                \dfrac{1}{n_{\mathrm{H}}}\dfrac{dn_{\mathrm{H}}}{dx} < \dfrac{2 \, R \, n_{\mathrm{H}}}{v_{\mathrm{IF}}}.
            \end{equation}
            This criterion represents the inverse of the characteristic length $L$ of variation of density and gives us the condition on the logarithmic derivative of the density to maintain a stationary state. It means that the criterion is equivalent to:
            \begin{equation}
                L > 5.5 \times 10^{-3} \, \left( \dfrac{10^{5}~ \mathrm{cm}^{-3}}{n_{\mathrm{H}}} \right) \left( \dfrac{v_{\mathrm{IF}}}{1 ~ \mathrm{km \, s}^{-1}} \right) ~ \mathrm{pc}.
            \end{equation}
            Thus, for dense gas, preexisting density variations on scales of a few mpc would be needed to invalidate our stationary hypothesis. In comparison, dense filaments have been argued to have a typical scale of 0.1 pc \citep{Arzoumanian2011}, so that in a wide range of environments, the acceleration or deceleration of the ionization front is slow enough for the stationary hypothesis to remain reasonable. However, smaller scale preexisting substructures might exist in some environments, in which the acceleration or deceleration of the ionization front would not permit the establishment of a stationary H/H$_2$ structure. Potential observational signatures of such nonstationary PDRs remain to be explored.
            
            Finally, our study adopts a 1D geometry. In 2D and 3D geometries, other structures can however emerge from instabilities induced by the propagation of the front itself. These instabilities have been proposed to explain the bright rims or elephant trunks structures. The linear analysis of thin-shell instabilities by \cite{Giuliani1979, Giuliani1980} were confirmed by numerical simulations by \cite{Garcia-Segura1996}. They also showed the importance of cooling in the neutral gas in the appearance of dynamic instabilities that lead to a rapid shell fragmentation. With the merging of the fragments, dense and massive clumps are created with a variety of shapes. \cite{Williams1999} showed that the clump cast a shadow that creates a tail of nonionized gas behind the clump, leading to further instabilities until the clump vanishes. In \cite{Whalen2008}, 3D radiation hydrodynamics simulations are presented to model the propagation of ionization fronts. They also concluded that the cooling efficiency in the shocked neutral medium plays a key role in the development of instabilities. Strong cooling leads to long clumpy structures whereas weak cooling leads to turbulent flows.
        
        \subsection{\label{sect_D_type_IF}D-type ionization fronts and critical IF velocity}
            While Eq. (\ref{eq_IF_velocity_F_EUV}) is necessarily true across a steady ionization front, the classical theory of ionization fronts (\citealt{Kahn1954}, \cite{Draine2011}, chap. 37) shows that a steady (plane-parallel) ionization front can only exist if the resulting velocity is either lower than $v_\mathrm{D}$ or higher than $v_\mathrm{R}$, with
            \begin{equation}
                v_{\mathrm{R}} = c_{\mathrm{II}} + \sqrt{c_{\mathrm{II}}^2 - c_{\mathrm{PDR}}^2 - \dfrac{v_{\mathrm{A}}^2}{2}},
                \label{eq_v_R}
            \end{equation}
            and
            \begin{equation}
                v_{\mathrm{D}} = c_{\mathrm{II}} - \sqrt{c_{\mathrm{II}}^2 - c_{\mathrm{PDR}}^2 - \dfrac{v_{\mathrm{A}}^2}{2}},
                \label{eq_v_D}
            \end{equation}
            where $c_{\mathrm{PDR}} = \sqrt{\gamma T/\mu}$ is the isothermal sound speed in the atomic region right after the ionization front with $\gamma$ the adiabatic index, T the PDR temperature, $\mu$ the mean mass per particle, $c_{\mathrm{II}}$ the isothermal sound speed of the H\,\textsc{ii} region and $v_{\mathrm{A}}$ is the Alfvén speed in the atomic region when a magnetic field is present.
            The Alfvén speed is simply taken as a free parameter here as we do not solve the magneto-hydrodynamics of the ionization front propagation.
            
            When the conditions ($F_\mathrm{EUV}$ and $n_{\mathrm{H}}$) are such that $v_{\mathrm{IF}} \ge v_{\mathrm{R}}$, the ionization front is called R-type. A R-type ionization front propagates supersonically relative to the neutral gas, and the gas mainly remains frozen in place as the IF propagates. This corresponds to very early phase of development of an H\,\textsc{ii} region, or to a later stage of acceleration of the front along a decreasing density ramp, as mentioned in Sect. \ref{sec_link_ratio_v_IF}. On the other hand, when the conditions are such that $v_{\mathrm{IF}} \le v_{\mathrm{D}}$, the ionization front is called D-type. Two types of D-type solutions are possible. Strong D-type fronts correspond to near pressure equilibrium between the ionized and neutral regions as can happen in the late stage of the H\,\textsc{ii} region expansion, approaching the static stage of a pressure-confined H\,\textsc{ii} region. In weak D-type fronts, the ionized gas is accelerated to roughly the ionized gas sound speed, with a corresponding pressure jump from the ionized to the neutral region. The atomic region thus has a higher pressure than the ionized region.
            When the conditions are such that we should have $ v_{\mathrm{D}} < v_{\mathrm{IF}} < v_{\mathrm{R}}$, no steady ionization front is possible. A shock front develops and precedes the ionization front, increasing the density of the neutral gas immediately before the ionization front so that $v_{\mathrm{IF}} = F_\mathrm{EUV}/n_\mathrm{H} \leq v_\mathrm{D}$.
            
            The values of $v_\mathrm{IF}$ that we can consider in our model are thus necessarily lower than $v_\mathrm{D}$ or higher than $v_\mathrm{R}$. 
            We consider a typical PDR temperature of $1000~\mathrm{K}$, giving $c_{\mathrm{PDR}} = 3.7~ \mathrm{km \,s}^{-1}$. The Alfvén speed in the PDR is taken in a wide range of $0$ to $5~ \mathrm{km \,s}^{-1}$, corresponding to a transverse magnetic field of 0 to a few 100 $\mu \mathrm{G}$. With $c_{\mathrm{II}} = 10~ \mathrm{km \,s}^{-1}$ we deduce that our upper boundary for $v_\mathrm{IF}$ in D-type fronts ranges between $0.7$ to $1.4~ \mathrm{km \,s}^{-1}$, while the lower boundary for $v_\mathrm{IF}$ in R-type fronts ranges between $18.6$ and $19.3~ \mathrm{km \,s}^{-1}$. Due to their high velocities relative to the neutral gas, R-type fronts can only exist in very short-lived phases and are thus most likely not relevant for our goal of characterizing the H\,\textsc{i} column density at the surface of dense molecular clouds in star forming regions where photoevaporation can be present. In situations where photoevaporation flows (also called photoabalation flows) are present, as in blister-type H\,\textsc{ii} regions, the fronts are expected to be close to D-critical, with $v_\mathrm{IF} \approx v_\mathrm{D}$ \citep{Henney2005, Henney2007}.
            
            Eq. (\ref{eq_v_R}) and Eq. (\ref{eq_v_D}) assume a transverse magnetic field. Ionization fronts with oblique magnetic fields have been studied in \cite{Williams2000} and \cite{Williams2001}. By rewriting the jump conditions for obliquely magnetized ionization fronts, they determined that a high transverse field shrinks the forbidden range of front velocities. Solutions of R and D-type ionization fronts and transitions between the regimes were investigated both from a mathematical perspective and physical relevance. They concluded that these obliquely magnetized ionization fronts may lead to transverse velocities and modulation of magnetic field, which we do not consider in this study.
            \subsection{Radiation pressure}
            Radiation pressure, caused by momentum transferred from the photons that are absorbed and scattered to the gas and dust, can be nonnegligible in H\,\textsc{ii} regions. \cite{Krumholz2009} studied the impact of the radiation pressure on the expansion of both embedded and blister H\,\textsc{ii} regions and concluded that radiation pressure could only become important for H\,\textsc{ii} regions that formed around very massive star clusters. However, their study assumed that all of the radiation momentum was given to the neutral shell directly, without considering the impact of the radiation pressure on the ionized gas. This impact on the ionized region has been exhaustively studied for static, pressure-confined H\,\textsc{ii} regions by \cite{Draine2011paper}, showing that it can cause the apparition of a central cavity with most of the ionized gas compressed into an ionized shell close to the ionization front. The impact of the radiation pressure on the expansion phase of spherical, embedded H\,\textsc{ii} regions has been investigated by \cite{Kim2016}, assuming the static stratification found by \cite{Draine2011paper} to remain valid during the expansion. They showed that the compression of the ionized gas in an outer ionized shell caused by radiation pressure can significantly affect the expansion of the H\,\textsc{ii} region due to the raised pressure in this ionized shell. Numerical simulations of the expansion of an embedded, spherical H\,\textsc{ii} region including radiative pressure on dust, gas-dust drift, and dust photodestruction (for the conditons of RCW 120) where presented by \cite{Akimkin2017}. They found that radiation pressure has a significant impact on the dynamics of expansion of the H\,\textsc{ii} region in the conditions of their model, and that dust drift can lead to the expulsion of a significant fraction of the dust from the H\,\textsc{ii} region (see also \citealt{Gustafsson2018}).
            
            While radiation pressure in embedded, spherical H\,\textsc{ii} region has thus been studied significantly in recent years, the impact of radiation pressure on photoevaporation flows (such as found in blister-type H\,\textsc{ii} regions) has however not been investigated in detail. \cite{Krumholz2009} consider the case of blister H\,\textsc{ii} region, but as they neglect the impact of radiation pressure on the ionized gas, they implicitly assume the photoevaporation flows to be unaffected by radiation pressure, and radiation pressure and photoevaporation to act independently on the neutral shell. \cite{Henney2007} argue that radiation pressure is not important for typical galactic blister H\,\textsc{ii} regions based on an order of magnitude comparison of the radiative pressure-induced acceleration to the acceleration in a photoevaporation flow. We assume here radiative pressure to be negligible for typical galactic dense PDRs found at the borders of blister H\,\textsc{ii} regions. In more extreme conditions, radiative pressure could be expected to quench photoevaporation by containing the flow of ionized gas. This could result in a decreased $v_\mathrm{IF}$, and thus a reduced impact on the H/H$_2$ transition as described by the results of our semi-analytical model (Sect. \ref{sec_analytic_overview} and \ref{sec_results}).
            

        \subsection{\label{subsect_observations}Comparison with observations}
            \begin{table*}
                \centering
                \scriptsize
                \caption{Physical parameters of a sample of PDRs illuminated by O stars.}
                \begin{tabular}{cccccccccc}
                    \hline
                    \hline
                    Source & $G_0$ & $n_{\mathrm{H}}$ & $n_{\mathrm{HII}}$ & $d_{\mathrm{star}}$ & Exciting star & $G_0/n_{\mathrm{H}}$ & $S_{\mathrm{EUV}}/S_{\mathrm{FUV}}$ & $\tau_{\mathrm{EUV}}$ & $v_{\mathrm{IF}}$\\
                     & [Habing] & $\left[\mathrm{cm}^{-3}\right]$ & $\left[\mathrm{cm}^{-3}\right]$ & [pc] & ($T_{eff}$, spec. type) & $\left[\mathrm{cm}^{3}\right]$ & & & $\left[\mathrm{km\,s}^{-1}\right]$ \\
                    \hline
                    California & 11[4-22] $^{(2)}$ & $10^{3}$-$10^{4}$ $^{(2)}$ & 0.1-5$\, \times \, 10^{1}$ $^{(3)}$ & 7.0 $^{(2)}$ & 37 000 K O7.5III $^{(2)}$ & $2.0\, \times \, 10^{-3}$ & 0.74 $^{(1)}$ & 0.89 & $9.3\, \times \, 10^{-3}$ \\
                    Horsehead & 35[18-64] $^{(2)}$ & $10^{4}$-$10^{5}$ $^{(2)}$ & 1-3.5$\, \times \, 10^{2}$ $^{(4)}$ & 3.5 $^{(5)}$& 33 000 K O9.5V $^{(2)}$ & $6.4\, \times \, 10^{-4}$ & 0.35 $^{(1)}$ & 3.99 & $9.5\, \times \, 10^{-4}$\\
                    Orion Bar & 20 000 $^{(6)}$ & $10^{5}$ $^{(6)}$ & $2.5\, \times \, 10^{3}$ $^{(7)}$ & 0.30 $^{(7)}$ & 39 000 K O5.5V-O6V $^{(7)}$ & 0.2 & 0.94 $^{(1)}$ & 3.80 & 0.79 \\
                    Carina & 10 000 $^{(8)}$ & $10^{5}$ $^{(8)}$ & $10^{2}$ $^{(9)}$ & 2.0 $^{(8)}$ & - - & 0.1 & $\sim$ 1 & 1.01 & 0.58 \\
                    S 106 & 1.6 $\times10^{5}$ $^{(10)}$ & 3.1 $\times 10^{5}$ $^{(10)}$ & 1.5-6$\times 10^{3}$ $^{(10,11)}$ & 0.44 $^{(10,12)}$ & O7-9 $^{(10)}$ & 0.52 & 0.38 $^{(1)}$ & 8.35 & 1.3 \\
                    M 17 & 1-8 $\times 10^{4}$ $^{(13)}$ & 2 $\times 10^{7}$ $^{(13)}$ & 700 $^{(14,15)}$ & 1.9 $^{(15)}$ & O4 $^{(16)}$ & $2.3\, \times \, 10^{-3}$ & 1.18 $^{(1)}$ & 6.73 & 1.5 $ \, \times \, 10^{-2}$ \\
                    IRAS 23133+6050 & 1.6 $\times10^{5}$ $^{(10)}$ & 3.1 $\times 10^{5}$ $^{(10)}$ & $2\, \times \, 10^{3}$ $^{(10,12)}$ & 0.1 $^{(10,17)}$ & O8.5-9.5 $^{(10)}$ & 0.52 & 0.24 $^{(1)}$ & 1 & 0.73 \\
                    \hline
                \end{tabular} \\
                {Note: $G_0$ is the FUV radiation field, $n_{\mathrm{H}}$ the PDR density, $n_{\mathrm{HII}}$ the H\,\textsc{ii} region density, $d_{\mathrm{star}}$ the distance between star and PDR, the stellar temperature and type, the $G_{0}/n_{\mathrm{H}}$ ratio, $S_{\mathrm{EUV}}/S_{\mathrm{FUV}}$ the star UV emission rate ratio, $\tau_{\mathrm{EUV}}$ the dust optical depth in the H\,\textsc{ii} region, and $v_{\mathrm{IF}}$ the deduced ionization front velocity.}
                
                {References: (1) \cite{Diaz-Miller1998}, (2) \cite{Habart2011}, (3) \cite{Boulanger1988}, (4) \cite{Compiegne2007}, (5) \cite{Schirmer2020}, (6) \cite{Joblin18}, (7) \cite{Salgado16}, (8) \cite{Wu18}, (9) \cite{Harper2009}, (10) \cite{Stock2015}, (11) \cite{Israel1978}, (12) \cite{Noel2005}, (13) \cite{Sheffer2013}, (14) \cite{Watson1981}, (15) \cite{Stutzki1988}, (16) \cite{Hoffmeister2008}, (17) \cite{Martin-Hernandez2003}.}
                \label{table_PDR_Observations}
            \end{table*}
            
            \begin{figure}
                \includegraphics[width=\hsize]{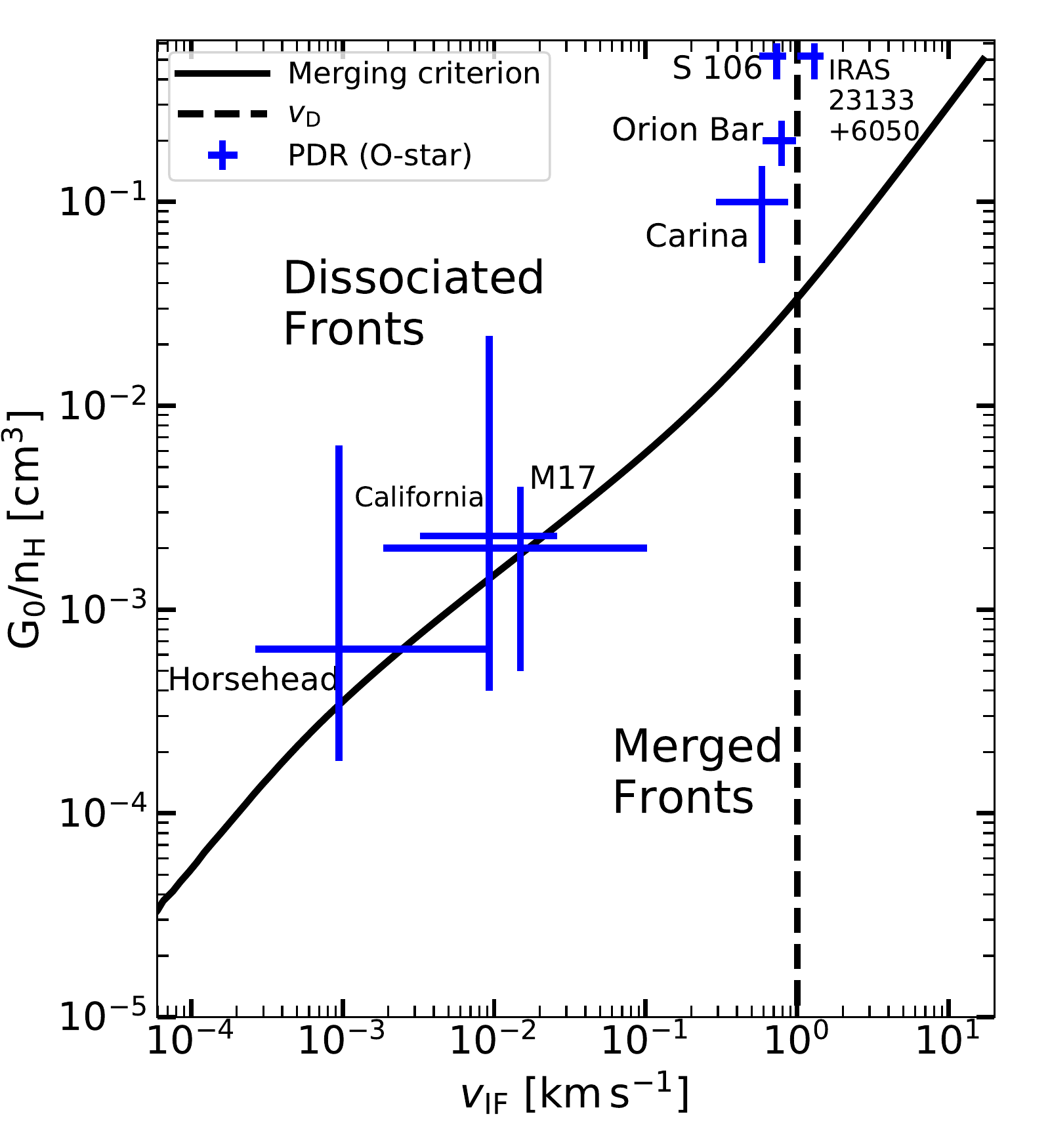}
                \caption{Comparison between the merging criterion from Fig. \ref{fig_Merging_criteria} and deduced parameters $G_{0}/n_{\mathrm{H}}$ and $v_{\mathrm{IF}}$ for observations. The vertical bar corresponds to the typical value $v_{\mathrm{D}} = 1~ \mathrm{km \,s}^{-1}$ deduced from Eq. (\ref{eq_v_D}).}
                \label{fig_Merging_criteria_Observations}
            \end{figure}
            
            Most observations of PDRs illuminated by B stars show that the H\,\textsc{ii} region is very close to the star and far from the dense PDR, so that photoevaporation is not relevant in that case. Furthermore \cite{Diaz-Miller1998} indicated that in B stars or lower mass cases, the H\,\textsc{ii} region is expected to be far closer to the star than the dissociation front. As the ionization and dissociation fronts are independent in this configuration, the semi-analytical model developed in this paper is no longer relevant.
            Still, if we do apply our semi-analytical model to the parameters of well-known B-star PDRs (NGC7023,...), it indeed yields a very small ionization front velocity, on the order of less than $10^{-4}~\mathrm{km\,s}^{-1}$ and a $G_{0}/n_{\mathrm{H}}$ ratio around $10^{-3}$-$10^{-2}~\mathrm{cm}^{3}$, leading to negligible effects of the dynamics. In the following comparison, we limit ourselves to PDRs illuminated by O stars.
            
            We present in Table \ref{table_PDR_Observations} the parameters relevant for our study to confront our conclusions to actual PDRs illuminated by O stars. For these PDRs, we deduced the column density between the star and the ionization front by multiplying the density estimated in the H\,\textsc{ii} region $n_{\mathrm{HII}}$ by the distance between the star and the PDR or the width of the region if mentioned. The ratio between the star emission rates of EUV and FUV photons are determined by making a log-linear interpolation on Table 1 from \cite{Diaz-Miller1998}. For Carina, there are a lot of stars illuminating the PDR, with a maximum temperature of 43 900 K \citep{Wu18}. We took the approximation of a ratio equal to 1 as expected for such a population of stars. Knowing $\tau_{\mathrm{EUV}}$ and $S_{\mathrm{EUV}}/S_{\mathrm{FUV}}$ for these objects, we use Eq. (\ref{eq_link_v_IF_ratio}) to calculate $v_{\mathrm{IF}}$. The result of the computation is given as the last column from Table \ref{table_PDR_Observations} and on Fig. \ref{fig_Merging_criteria_Observations} with error bars. We can see that for all these objects, the dynamic effects have to be important, as they are close to the merging criterion. For the Horsehead and California, the error bars even spread into the merged fronts space.
            
            The PDRs seems to form two groups in the $v_{\mathrm{IF}}$ versus $G_{0}/n_{\mathrm{H}}$ space, although whether this distinction is an artifact of our small sample or a physical effect remains to be confirmed.
            For the high-excitation PDRs (high $G_{0}/n_{\mathrm{H}}$) in our sample (Orion Bar, Carina, S 106, and IRAS 23133+6050), our estimated values of $v_\mathrm{IF}$ seem to show a clear saturation of the ionization front velocity at the expected value for D-critical fronts (indicated by the dashed vertical line), as predicted by ionization front theory (see Sect. \ref{sect_D_type_IF}). Such D-critical fronts are indeed expected at the dense walls of blister H\,\textsc{ii} regions \citep{Henney2005,Henney2007}. This group of PDRs clearly falls in the nonmerged fronts regime, which is compatible with ALMA observations of the Orion Bar showing a relatively extended atomic region separating the ionization front and the H/H$_2$ transition \citep{Goicoechea16}. As found in Sect. \ref{sec_advection_effect}, we nevertheless predict moderate but non negligible effects of the advection dynamics on the H/H$_2$ transition. 
            A second group at low $G_{0}/n_{\mathrm{H}}$ ratio is composed of the Horsehead nebula, California, and M17. The ionization front velocity estimates for these PDRs seems to be close to the merging boundary, although the large error bars do not allow to conclude whether they fall in the merged fronts regime or in the nonmerged front regime. The velocities found correspond to noncritical weak D-type ionization fronts. If the line-up of these PDRs on the merging boundary is confirmed by more accurate estimations, it might indicate a specific dynamics of merged fronts causing a saturation at the critical merging velocity. Based on these results, we thus expect strong effects of the advection dynamics with merged, or close to merged, ionization front and dissociation front in these low excitation PDRs illuminated by O stars. Strong effect on the emission lines of H$_2$ could in particular be expected in these low $G_{0}/n_{\mathrm{H}}$ PDRs, in contrast to the conclusions of \cite{Storzer98} who focused on high $G_{0}/n_{\mathrm{H}}$ conditions representative of Orion Bar-like PDRs.
            
        \subsection{\label{subsect_Hydra}Comparison to Hydra PDR Code simulations}
\begin{figure}
                \includegraphics[width=\hsize]{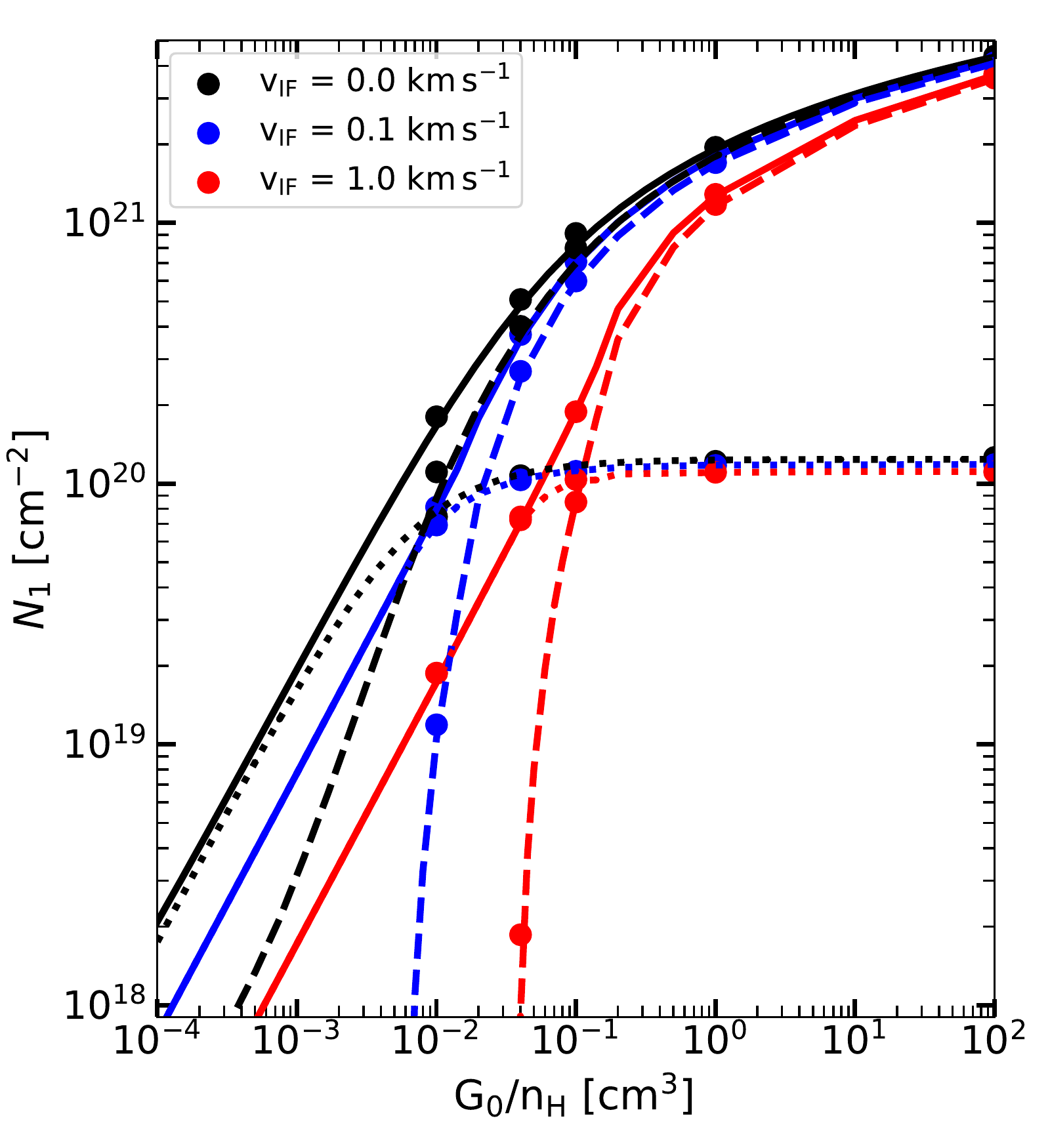}
                \caption{$N_{1}$ integrated over the atomic region (dashed lines), over the molecular region (dotted lines), and over the whole PDR, $N_{1,\mathrm{tot}}$ (solid lines) as functions of the $G_0/n_{\mathrm{H}}$ ratio. The different colors (black, blue, and red) represent $v_{\mathrm{IF}} = 0, 0.1~\mathrm{and}~1~ \mathrm{km \,s}^{-1}$. The different points are obtained with the Hydra PDR Code.}
                \label{fig_N1_static_vs_dynamical_vs_HYDRA}
            \end{figure}

            In order to compare our $N_{1,\mathrm{tot}}$ semi-analytical predictions to Hydra PDR Code simulations, we implemented an advection of the cloud through a punctual ionization front. We plot in Fig. \ref{fig_N1_static_vs_dynamical_vs_HYDRA} the predictions of $N_{1,\mathrm{tot}}$ and $N_1$ integrated over the atomic and molecular regions for a number of simulations as circles. The agreement is very good with a maximum deviation of a factor 1.2 for every point but one. The latter is for $N_1$ integrated over the atomic region for $v_{\mathrm{IF}} = 1~\mathrm{km\,s}^{-1}$ and $G_{0}/n_{\mathrm{H}} = 0.04$. It has a deviation of 2, which can easily be explained by the lack of resolution of the Hydra PDR Code simulation. Because we examine a case very close to front merging, the resolution between the ionization and photodissociation fronts may be insufficient for further precision, yet it is in excellent agreement with the semi-analytical prediction.
            
    \section{\label{sec_conclusions}Conclusions}
        To sum up, we built a semi-analytical model of the H/H$_2$ transition in a 1D plane-parallel PDR illuminated by FUV radiation with advancing ionization front resulting from the photoevaporation mechanism. We investigated the stationary equilibrium state in the ionization front reference frame where its propagation translates into advection of the neutral gas through the front. We did not solve the hydrodynamics of the gas through the front, but included the advection term to account for the impact of the steady flow. The advection positively adds up to H$_2$ formation on dust and its photodissociation by FUV photons. Our radiative transfer takes into account not only the extinction by dust but also the H$_2$ self-shielding.
        
        We provide a simple set of analytical expressions allowing to compute the total H\,{\textsc i} column density $N_{1,\mathrm{tot}}$ in the presence of photoevaporation: Knowing $G_0$, $n_{\mathrm{H}}$, the metallicity $Z'$, the stellar type of the illuminating star and $N_{\mathrm{HII}}$ the total proton column density in the H\,\textsc{ii} region, one can first estimate the velocity of the ionization front $v_{\mathrm{IF}}$ from Eq. (\ref{eq_link_v_IF_ratio}). One then has to find whether the fronts are merged or not by comparing the $G_0/n_{\mathrm{H}}$ ratio to the critical value found in Eq. (\ref{eq_fit_merging_criterion_metallicity}). Eq. (\ref{eq_N_1_tot_static}) then gives the static prediction value of $N_{1,\mathrm{tot}}$, and Eq. (\ref{eq_shift_N1_tot_metallicity}) finally gives the desired dynamical prediction of $N_{1,\mathrm{tot}}$.
        
         We present a summary of our results:
        \begin{itemize}
            \item[$\bullet$] The advection of the gas through the ionization front induces a change in the location of the H/H$_2$ transition, which is closer to the ionization front compared to the static case. This effect reduces the width of the atomic region.
            \item[$\bullet$] If the ionization front velocity exceeds a certain value, the atomic region disappears. The photodissociation front and the ionization front then form a merged structure. The merging of the fronts has been studied in previous papers \citep{Bertoldi96, Storzer98}. We provide a new, more accurate criterion to discriminate whether or not the fronts are merged.
            \item[$\bullet$] The total atomic hydrogen column density $N_{1,\mathrm{tot}}$ integrated in the PDR is decreased in the presence of a propagating ionization front. When the fronts are not merged, the absolute difference in $N_{1,\mathrm{tot}}$ between the static case and the case of a propagating ionization front depends only on the velocity of the ionization front. When the fronts are merged, it is the relative difference that depends only on the velocity of the ionization front. We provide analytical approximations in both cases.
            \item[$\bullet$] The metallicity plays a key role in this study. A lower metallicity changes the merging criterion. Indeed, in such a case, a higher ionization front velocity is needed to merge the fronts. It also changes the magnitude of the total atomic hydrogen column density. The two-regimes dichotomy is still valid with different metallicities.
            \item[$\bullet$] Observations of PDRs illuminated by O-stars are close to the merging criteria, where the dynamical effects are strong. For this kind of objects, we expect $N_{1,\mathrm{tot}}$ to be much lower in the advection approach than with a static one, especially for low excitation PDRs.
        \end{itemize}
        
        By taking into account the photoevaporation, we found that the H/H$_2$ transition is closer to the ionization front than modeled by previous static studies. Some H$_2$ is then found in a hotter region, richer in FUV photons. One can then expect H$_2$ ro-vibrational lines to be more intense as collisions and UV-pumping are more efficient. As a lot of H$_2$ ro-vibrational lines will be observed by the upcoming JWST, taking into account the dynamical effects highlighted in this paper will be essential for the modeling of observations. The effects are expected to cause larger differences for low excitation PDRs, such as the Horsehead Nebulae.


    \begin{acknowledgements}
        We thank the anonymous referee for helpful comments, conversations, and suggestions. This work was supported by the Programme National “Physique et Chimie du Milieu Interstellaire” (PCMI) of CNRS/INSU with INC/INP co-funded by CEA and CNES.
        
    \end{acknowledgements}

    \bibliographystyle{aa} 
    \bibliography{biblio}

    \appendix
        \section{\label{Appendix_Analytic_solution}Solving Eq. (\ref{eq_static_ODE}) with Eq. (\ref{eq_36_H2_self_shielding}) as H$_2$ self-shielding function}
            Let us find the position of the H/H$_2$ transition defined as the position where $x_1 = x_2$, and, as a reminder, $x_{1,2}$ = $d \tau_{1,2}/d \tau_g$.
            
            Writing $x_1^0$, $x_2^0$, $\tau_1^0$, $\tau_2^0$, $\tau_{g}^0$ the values of abundances and optical depths at the position of the transition and by using the definition of the transition and $1 = x_1^0 + 2x_2^0$, we can deduce that $x_2^0 = 1/3$. We can then rewrite Eq. (\ref{eq_static_ODE}) at the position of the transition as follows:
            \begin{equation}
                \dfrac{1}{3} = \dfrac{1}{2 + \dfrac{\mathcal{P}_0}{R} \, \dfrac{G_0}{n_{\mathrm{H}}} \, \exp{\left(-\tau_g^0\right)}\,f_{\mathrm{shield}}(\tau_2^0/\sigma_g)}.
            \end{equation}
            \begin{equation}
                \tau_g^0 = \tau_1^0 + 2 \tau_2^0 = \ln{\left[\dfrac{\mathcal{P}_0}{R} \dfrac{G_0}{n_{\mathrm{H}}}{f_{\mathrm{shield}}(\tau_2^0/\sigma_g)}\right]}.
                \label{eq_1_analyt_syst_transit_pos}
            \end{equation}
            
            We also rewrite Eq. (\ref{eq_static_ODE}) at any position (and not just at the transition):
            \begin{equation}
                \dfrac{d \tau_g}{d \tau_2} = 2 + \dfrac{\mathcal{P}_0}{R} \, \dfrac{G_0}{n_{\mathrm{H}}} \, \exp{\left(-\tau_g\right)}\,f_{\mathrm{shield}}(\tau_2/\sigma_g).
            \end{equation}
            Using that:
            \begin{equation}
                \dfrac{d \tau_g}{d \tau_2} = \dfrac{d \left( \tau_1 + 2 \tau_2 \right)}{d \tau_2} = \dfrac{d \tau_1}{d \tau_2} + 2,
            \end{equation}
            we deduce that :
            \begin{equation}
                 \dfrac{d \tau_1}{d \tau_2} = \dfrac{\mathcal{P}_0}{R} \, \dfrac{G_0}{n_{\mathrm{H}}} \, \exp{\left(-\tau_g\right)}\,f_{\mathrm{shield}}(\tau_2/\sigma_g).
            \end{equation}
            
            \begin{equation}
                \dfrac{d\tau_2}{d\tau_1} = \left( \dfrac{\mathcal{P}_0}{R} \dfrac{G_0}{n_{\mathrm{H}}}f_{\mathrm{shield}}(\tau_2/\sigma_g) e^{-(\tau_1+2\tau_2)} \right)^{-1}.
                \label{eq_appendix_dtau2_dtau1}
            \end{equation}
            
            To solve this, we use Eq. (\ref{eq_36_H2_self_shielding}) as the H$_2$ self-shielding function. This function is composed of two sections. One for low values of H$_2$ column densities with $N_{2} < 10^{14} ~ \mathrm{cm}^{-2}$, and one for $N_{2} > 10^{14} ~ \mathrm{cm}^{-2}$. Solving Eq. (\ref{eq_appendix_dtau2_dtau1}) consequently requires to first solve it for the first case, giving us a solution from $\tau_2 = 0$ to $\tau_2 = \tau_2^{\mathrm{shield}} = \sigma_g \times \left(10^{14} ~ \mathrm{cm}^{-2} \right) = 1.9 \times 10^{-7}$. After that, we solve it for the second case, from $\tau_2 = \tau_2^{\mathrm{shield}}$ to $\tau_2 \rightarrow + \infty$. For the first case where the H$_2$ self-shielding is equal to 1, Eq. (\ref{eq_appendix_dtau2_dtau1}) can be rewritten as:
             \begin{equation}
                \dfrac{d\tau_1}{d\tau_2} = \dfrac{\mathcal{P}_0}{R} \dfrac{G_0}{n_{\mathrm{H}}} e^{-(\tau_1+2\tau_2)},
            \end{equation}
            with the following solution giving $\tau_1$ as a function of $\tau_2$:
            \begin{equation}
                \tau_1(\tau_2) = \ln{\left(1 + \dfrac{1}{2}\dfrac{\mathcal{P}_0}{R} \dfrac{G_0}{n_{\mathrm{H}}} e^{-2\tau_2}\right)}.
            \end{equation}
            The boundary where $\tau_2 = \tau_2^{\mathrm{shield}}$ gives:
            \begin{equation}
                \tau_1\left(\tau_2^{\mathrm{shield}}\right) = \ln{\left(1 + \dfrac{1}{2}\dfrac{\mathcal{P}_0}{R} \dfrac{G_0}{n_{\mathrm{H}}} e^{-2 \tau_2^{\mathrm{shield}}}\right)}.
            \end{equation}
            In the other case, where the self-shielding of H$_2$ is a decreasing power-law, Eq. (\ref{eq_appendix_dtau2_dtau1}) is rewritten as:
            \begin{equation}
                \dfrac{d\tau_2}{d\tau_1} = \left( \dfrac{\mathcal{P}_0}{R} \dfrac{G_0}{n_{\mathrm{H}}}\left(\dfrac{\tau_2}{\tau_2^{\mathrm{shield}}}\right)^{-3/4} e^{-(\tau_1+2\tau_2)} \right)^{-1},
            \end{equation}
           
            with the following solution:
            \begin{multline}
                \tau_2\left( \tau_1\right) = \dfrac{1}{2} \Gamma ^{-1} \left[ \dfrac{1}{4},\Gamma \left( \dfrac{1}{4},\tau_2^{\mathrm{shield}} \right) \right. \\
                \left.+ \left(1 + \dfrac{1}{2} \dfrac{\mathcal{P}_0}{R} \dfrac{G_0}{n_{\mathrm{H}}} \exp{\left( 2\tau_2^{\mathrm{shield}}\right)} - e^{\tau_1}\right)/\xi \right],
                \label{eq_full_solution}
            \end{multline}
            
            with $\Gamma$ the incomplete Gamma function so that $\dfrac{\partial \Gamma(a,x)}{\partial x}=-x^{a-1}\times e^{-x}$, $\Gamma^{-1}$ the inverse incomplete Gamma function and:
            
            \begin{equation}
                \xi = \dfrac{\left(\tau_2^{\mathrm{shield}}\right)^{3/4}}{\sqrt[4]{2}} \dfrac{\mathcal{P}_0}{R} \dfrac{G_0}{n_{\mathrm{H}}}.
                \label{eq_xi_analytic_cst}
            \end{equation}
            The $\tau_2^{\mathrm{shield}}$ parameter is very small compared to typical values of $\tau_1$ or $\tau_2$ at the H/H$_2$ transition. We can thus neglect the exponential term involving $\tau_2^{\mathrm{shield}}$ in Eq. \ref{eq_full_solution}, so that:
            \begin{equation}
                \tau_2\left( \tau_1\right) = \dfrac{1}{2} \Gamma ^{-1} \left[ \dfrac{1}{4},\Gamma \left( \dfrac{1}{4}, \tau_2^{\mathrm{shield}} \right) + (1- e^{\tau_1})/\xi \right].
                \label{eq_2_analyt_syst_transit_pos}
            \end{equation}
            To finally obtain $\tau_1$ at the position of the transition we need to solve the system given by Eq. (\ref{eq_1_analyt_syst_transit_pos}) (valid only at the position of the transition) and Eq. (\ref{eq_2_analyt_syst_transit_pos}) (valid everywhere). We solved that system numerically to obtain the results shown in Figure \ref{fig_Analytic_vs_Sternberg}.
            
            Moreover, as we are interested in the total atomic hydrogen column density $N_{1,\mathrm{tot}}$, we derive here Eq. (\ref{eq_N_1_tot_static}). Expressing Eq. (\ref{eq_2_analyt_syst_transit_pos}) as $\tau_1(\tau_2)$:
            \begin{equation}
                \tau_1(\tau_2) = \ln{\left(1 + \xi \, \left(\Gamma{\left( \dfrac{1}{4}, \tau_2^{\mathrm{shield}} \right)} - \Gamma{\left( \dfrac{1}{4}, 2 \, \tau_2 \right)} \right) \right)},
                \label{eq_tau_1_static}
            \end{equation}
            from which we can derive $N_{1,\mathrm{tot}}$ by taking the limit $\tau_{2} \rightarrow +\infty$. Indeed, the fraction of atomic hydrogen is decreasing as $\tau_{\mathrm{g}} \rightarrow +\infty$, so at infinity the gas is strictly in molecular form, $\tau_{2}$ or $\tau_{\mathrm{g}}$ is the same in this limit. In the end, we obtain:
            \begin{multline}
                N_{1,\mathrm{tot}} = \dfrac{1}{\sigma_g}\ln{\left(1 + \xi \,\Gamma{\left( \dfrac{1}{4}, \tau_2^{\mathrm{shield}} \right)} \right)} \\
                = \dfrac{1}{\sigma_g} \ln{\left(1 + 2.97 \, \left(\sigma_g \times 10^{14}~\mathrm{cm}^{-2} \right)^{3/4} \dfrac{\mathcal{P}_0}{R} \dfrac{G_0}{n_{\mathrm{H}}} \right)}.
                \label{eq_N_1_tot_static_appendix}
            \end{multline}
        
        \section{\label{Appendix_C_extinction}C\textsc{i} extinction}
            Far-UV photons are susceptible of being absorbed by atomic carbon C\,\textsc{i} in the continuum, thus possibly contributing to the shielding of H$_2$. The resulting opacity was neglected in this paper, and we show in this appendix that while extinction by gas phase carbon can be important for the global structure of a PDR, its impact on the H/H$_2$ transition is negligible.
            
            In the FUV band, the atomic carbon absorbs from 912 \AA~to 1101 \AA~with an almost constant absorption cross-section $\sigma_{\mathrm{C}} \simeq 1.7 \times 10^{-17} \, \mathrm{cm}^{2}$ \citep{Nahar1997}. In this band and at a given position, it diminishes the flux by a factor of $\exp{\left(-\sigma_{\mathrm{C}} N_{\mathrm{C}}\right)}$, with $N_{\mathrm{C}}$ the column density of C\,\textsc{i} integrated from the edge of the cloud to the given position. Adding its effect to the semi-analytical model of this paper would thus require taking into account a more extensive chemical network in order to compute the column density of atomic carbon. The hypothesis maximizing this extinction would be to assume that all the carbon is in atomic form. With an abundance around $x(\mathrm{C}) = 10^{-4}$, one can rewrite the extinction term as $\exp{\left(-\sigma_{\mathrm{C}} N_{\mathrm{C}}\right)} = \exp{\left(-\sigma_{\mathrm{C}} \, x(\mathrm{C}) \, N\right)}$ with $\sigma_{\mathrm{C}} x(\mathrm{C}) \simeq 1.7 \times 10^{-21} \, \mathrm{cm}^{2}$, which is very close to dust extinction cross-section $\sigma_g = 1.9 \times 10^{-21} \, \mathrm{cm}^{2}$. So, at best, atomic carbon extinction would be equal to dust extinction. \cite{Sternberg14} showed that the conditions for which dust extinction dominates over H$_2$ self-shielding are met in the strong field limit. In the weak field limit, where $G_0/n_{\mathrm{H}} \leq 10^{-2} \, \mathrm{cm}^{3}$, the H/H$_2$ transition is driven by H$_2$ self-shielding and not by dust, and atomic carbon extinction can thus be expected to have no impact on the H/H$_2$ transition. In the strong field limit ($G_0/n_{\mathrm{H}} \geq 10^{-2}\, \mathrm{cm}^{3}$), its impact depends on the abundance of atomic carbon before the H/H$_2$ transition.
            
\begin{figure}
                \includegraphics[width=\hsize]{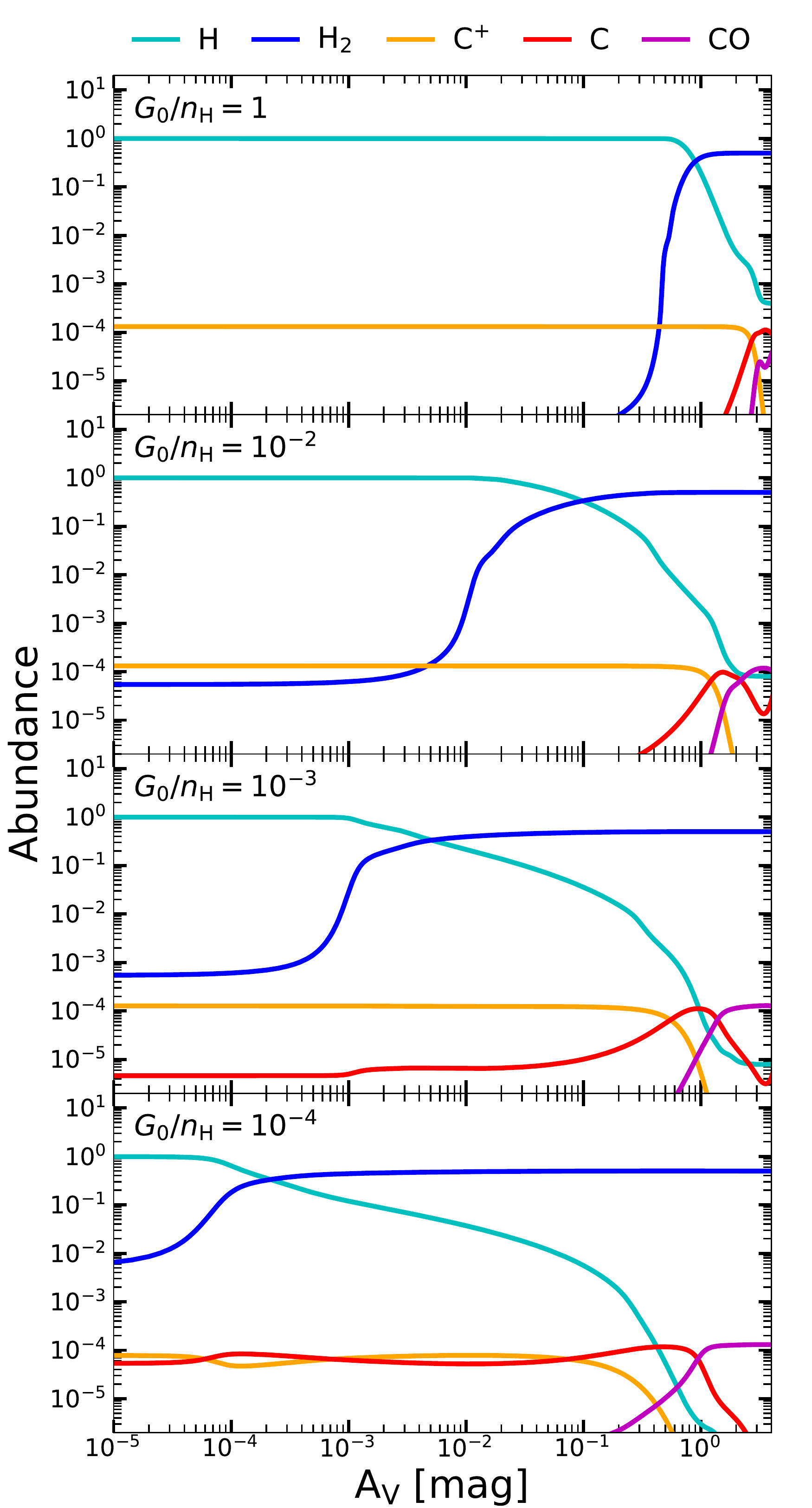}
                \caption{Abundance profiles of H in cyan, H$_2$ in blue, C$^{+}$ in orange, C in red, and CO in violet as functions of A$_{\mathrm{V}}$ for models with $G_0/n_{\mathrm{H}} = 1$, $10^{-2}$, $10^{-3}$, and $10^{-4} \, \mathrm{cm}^{3}$ obtained using Meudon PDR Code.}
                \label{Appendix_C_extinction_profiles}
            \end{figure}

\begin{figure}
                \includegraphics[width=\hsize]{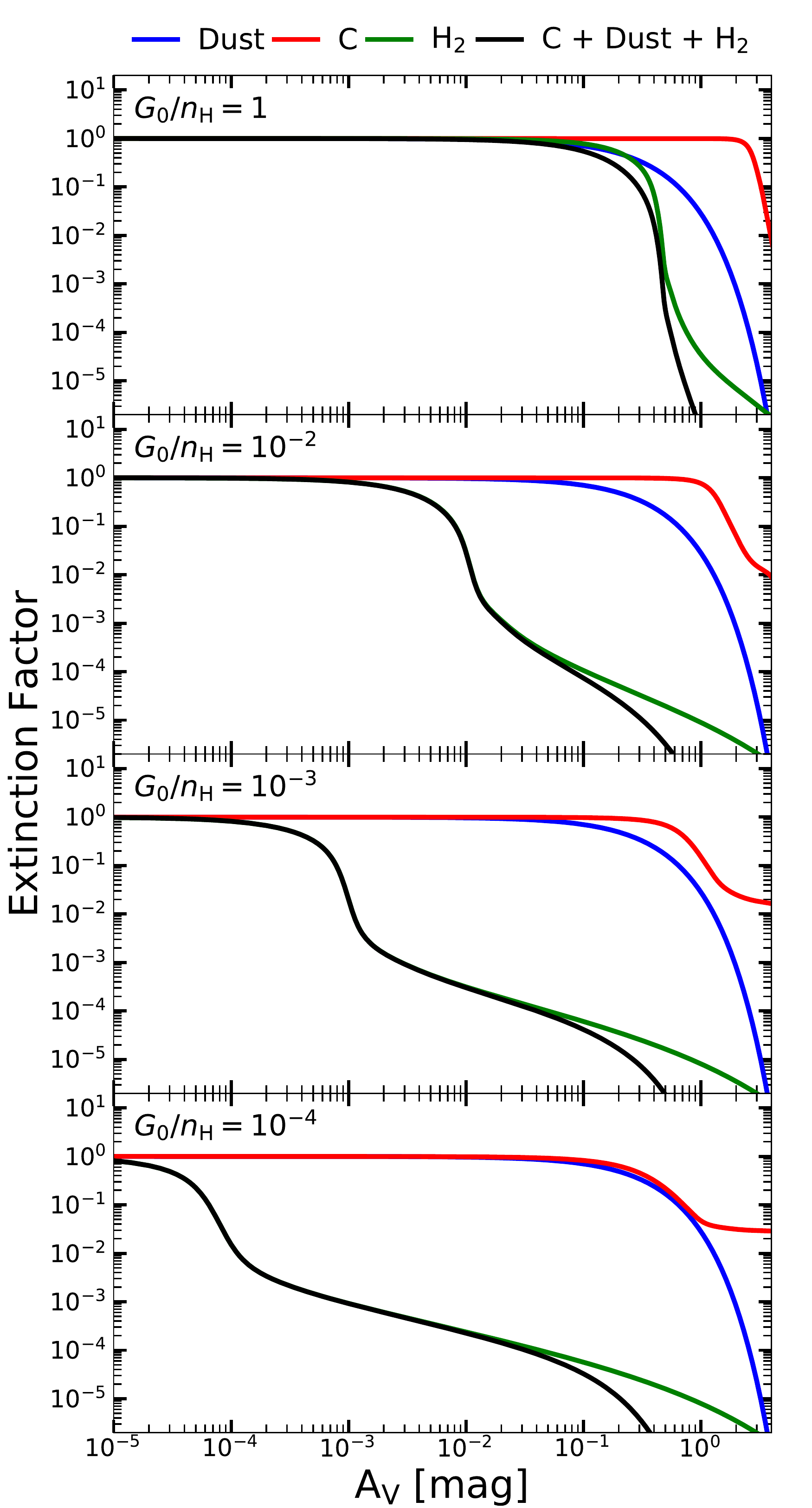}
                \caption{Extinction factor of dust in blue, of C in red, of H$_2$ self-shielding in green, and of all of them combined in black, as functions of $A_{\mathrm{V}}$. C\,\textsc{i} and H$_2$ column densities are taken from Meudon PDR Code simulations for models with $G_0/n_{\mathrm{H}} = 1$, $10^{-2}$, $10^{-3}$, and $10^{-4} \, \mathrm{cm}^{3}$.}
                \label{Appendix_C_extinction_vs_dust_H2}
            \end{figure}

            To investigate this impact, we used the Meudon PDR Code. The Meudon PDR Code \citep{LePetit2006} simulates a stationary, plane-parallel PDR, computing self-consistently the radiative transfer both in the continuum and in lines \citep{Goicoechea07}, the chemistry with an exhaustive chemical network and the thermal balance of the gas. We obtained the results presented in Fig. \ref{Appendix_C_extinction_profiles} and \ref{Appendix_C_extinction_vs_dust_H2} for different $G_0/n_{\mathrm{H}}$ ratio, from $1$ to $10^{-4} \, \mathrm{cm}^{3}$. On Fig. \ref{Appendix_C_extinction_profiles}, we have the abundances profiles of H, H$_2$, C$^{+}$, C, and CO as functions of A$_{\mathrm{V}}$. Fig. \ref{Appendix_C_extinction_vs_dust_H2} presents the extinction terms of dust, C, H$_2$, and the combination of the three as functions of A$_{\mathrm{V}}$. For dust and carbon, the extinction terms are $\mathrm{exp}(-\sigma_g \,\, N)$ and $\mathrm{exp}(-\sigma_\mathrm{C} \,\, x(\mathrm{C}) \,\, N)$ respectively and, for H$_2$, it is equal to the shielding function $f_{shield}(\mathrm{H}_2)$. These extinction terms are not extracted directly from the PDR models because the PDR code computes the sum of extinctions at each wavelength and it is not convenient to separate the various contributions. Instead, since here we just need simple estimations of these extinction terms, we computed them with the above expressions, and for the carbon extinction term, we use the abundance of C computed at each position by the code.
            
            In the high $G_0/n_{\mathrm{H}}$ regime (see e.g., $G_0/n_{\mathrm{H}} = 1 \, \mathrm{cm}^{3}$, top panel), atomic carbon is not very abundant at the H/H$_2$ transition position with $x(\mathrm{C}) \simeq 10^{-7}$ for $G_0/n_{\mathrm{H}} = 1 \, \mathrm{cm}^{3}$, so that extinction due to C\,\textsc{i} is smaller than dust extinction. When we compare the actual extinction factors in the top panel of Fig. \ref{Appendix_C_extinction_vs_dust_H2}, we see that indeed, C\,\textsc{i} extinction is negligible until around $A_{\mathrm{V}} \simeq 2.3$, deeper than the transition, found around $A_{\mathrm{V}} = 1$. As $G_0/n_{\mathrm{H}}$ is decreased, the abundance of atomic carbon in the atomic region increases. In the weak field case, dust and atomic carbon extinctions are very close because carbon is almost entirely in neutral form. But as expected for the weak field regime \citep{Sternberg14}, one can see that from the edge of the cloud to the H/H$_2$ transition, the column density is far too small for the extinction factor (both by dust and by atomic carbon) to significantly differ from 1. FUV extinction is mainly driven by H$_2$ self-shielding, with negligible contributions of carbon or dust extinction. In intermediate cases, like panels 2 and 3, dust slowly takes a leading importance, but extinction due to C\,\textsc{i} always remains negligible up to the H/H$_2$ transition. Deeper in the cloud, atomic carbon continuum extinction of course becomes important, and in particular, we see that it cannot be neglected for the transition to CO, but this region is not considered in this paper.
            
            In conclusion, we find atomic carbon extinction to be negligible in all cases for the purpose of the study of the H/H$_2$ transition. In the case of high $G_0/n_{\mathrm{H}}$, the very low fraction of atomic carbon in the atomic region makes it negligible compared to dust extinction. In the case of low $G_0/n_{\mathrm{H}}$, where the fraction of atomic carbon in the atomic region increases to significant levels, the H/H$_2$ transition is dominated by H$_2$ self-shielding and appears at a very low column density, so that extinction by both dust and atomic carbon is negligible.
                
\end{document}